\documentclass[aps,prl,superscriptaddress,twocolumn,10pt]{revtex4-2}

\usepackage{amssymb}
\usepackage{cancel}
\usepackage{color,graphicx}
\usepackage{amsmath}
\usepackage{amsbsy}
\usepackage{amsthm}
\usepackage{bbm}
\usepackage{epsfig}
\usepackage{float}
\usepackage{graphicx}
\usepackage{subfigure}
\usepackage{dcolumn}
\usepackage{bbm}
\usepackage{color,epstopdf}
\usepackage{amscd}
\usepackage{amsfonts}  
\usepackage[]{amsmath}
\usepackage{amssymb}    
\usepackage{mathrsfs}
\usepackage{verbatim}
\usepackage[]{cases}
\usepackage{amsmath}
\usepackage{wasysym}
\usepackage[latin9]{inputenc}
\usepackage[T1]{fontenc}
\usepackage{mathtools}
\usepackage{todonotes}
\usepackage[normalem]{ulem}
\usepackage[colorlinks=true,linkcolor=blue,urlcolor=blue,citecolor=blue,pdfusetitle]{hyperref}
\usepackage{cleveref}
\usepackage{libertine}
\usepackage{hyperref}

\definecolor{ao}{rgb}{0.0, 0.5, 0.0}
\definecolor{mypink}{rgb}{0.858, 0.188, 0.478}
\definecolor{mygreen}{rgb}{0.0, 0.5, 0.0}

\begin{document}

\title{The role of quantum coherence in energy fluctuations}

\author{S. Gherardini}\thanks{These authors contributed equally to this work}
\affiliation{CNR-INO \& LENS, via G.  Sansone 1, I-50019 Sesto Fiorentino, Italy.}
\affiliation{\mbox{Department of Physics and Astronomy, University of Florence,} via G.  Sansone 1, I-50019 Sesto Fiorentino, Italy.}
\affiliation{CNR-IOM DEMOCRITOS Simulation Center and SISSA, Via Bonomea 265, I-34136 Trieste, Italy}
\author{A. Belenchia}\thanks{These authors contributed equally to this work}
\affiliation{Institut f\"{u}r Theoretische Physik, Eberhard-Karls-Universit\"{a}t T\"{u}bingen, 72076 T\"{u}bingen, Germany}
\affiliation{Centre for Theoretical Atomic, Molecular and Optical Physics, School of Mathematics and Physics, Queen's University Belfast, Belfast BT7 1NN, United Kingdom}
\author{M. Paternostro}
\affiliation{Centre for Theoretical Atomic, Molecular and Optical Physics, School of Mathematics and Physics, Queen's University Belfast, Belfast BT7 1NN, United Kingdom}
\author{A. Trombettoni}
\affiliation{Department of Physics, University of Trieste, Strada Costiera 11, I-34151 Trieste, Italy}
\affiliation{CNR-IOM DEMOCRITOS Simulation Center and SISSA, Via Bonomea 265, I-34136 Trieste, Italy}

\begin{abstract}
We discuss the role of quantum coherence in the energy fluctuations of open quantum systems. To this aim, we introduce a protocol, to which we refer to as the end-point-measurement scheme, allowing to define the statistics of energy changes as a function of energy measurements performed only after {the evolution of the initial state.}
At the price of an additional uncertainty on the initial energies, this approach prevents the loss of initial quantum coherences and enables the 
estimation of their effects on energy fluctuations. We demonstrate our findings by running an experiment on the IBM Quantum Experience superconducting qubit platform.
\end{abstract}

\maketitle

When the size of a physical system is scaled down to the micro-/nano-scopic domain, fluctuations of relevant quantities start playing a pivotal role in establishing the energetics of the system. Such fluctuations obey fundamental relations, known as \textit{fluctuation theorems}, that recast
the laws of thermodynamics in such a new regime. Should the range of energies involved in a given system bring
its dynamics within the domain of quantum theory, the very nature of {energy} fluctuations become even more interesting as encompassing both classical (i.e. thermal) and quantum contributions. The characterization of the latter, and the understanding of how they conjure with the former to set the dynamics of fundamental energy transformations, are very stimulating open problems. 

One of the key achievements of the field of thermodynamics of quantum processes~\cite{Vinjanampathy2016,Sagawa2014,BookQT,deffner2019qtd} is the identification of a strategy for the assessment of the
energetics stemming from non-equilibrium quantum dynamics.
The so-called two-point measurement (TPM)
protocol~\cite{TalknerPRE2007,EspositoRMP2009,CampisiRMP2011}, where the energy is measured both at the initial and final time, has been introduced to determine the work statistics of a quantum system driven by a time-dependent protocol. 
However, in quantum mechanics, measurements condition the evolution of {the measured} system~\cite{BookJacobs}. In particular, in TPM an energy measurement performed before the dynamics takes place destroys the quantum coherences in the initial state of the system, forcing it into an energy eigenstate~\cite{AllahverdyanPRE2014,LostaglioNATCOMM2015}. Such a loss of coherence is common to interferometric formulations of the TPM protocol, which have been put forward to ease the inference of the energetics of out-of-equilibrium systems~\cite{Mazzola2013,Dorner2013,Batalhao2014}.

Recently, much effort has been devoted to understand the role of coherence in quantum thermodynamics~\cite{SolinasPRE2015,SolinasPRA2016,AlhambraPRX2016,AbergPRX2018,LostaglioPRL2018,XuPRA2018,FrancicaPRE2019,SantosnpjQI2019,MingoQuantum2019,PhysRevX.9.031029,MicadeiPRL2020}. In particular in
Refs.~\cite{SolinasPRE2015,SolinasPRA2016,XuPRA2018,LevyArxiv2019}
full counting statistics~\cite{NazarovEURPHYS2003,ClerkPRA2011}
has been used to study work fluctuations in quantum systems initialized in an arbitrary state, pointing out that the quantum interference stemming from considering quantum coherences
could lead to negative quasi-probability work distributions~\cite{HoferPRL2016}.

In this paper, we propose an \textit{end-point-measurement (EPM)} protocol to quantify the statistics of energy-change fluctuations
in the (possible) presence of quantum coherence in the initial
state of a system. Such a protocol removes the need for the first projective measurement required by TPM, thus preventing the collapse of the initial state of the system onto the energy basis. This is in contrast with recent proposals such as Ref.~\cite{MicadeiPRL2020}, where the system has to be prepared in a mixture of eigenstates of an observable $O$ that does not commute with the Hamiltonian of the system. This is equivalent to an experiment measuring $O$ at the initial time so that in each trajectory the starting point is an eigenstate of $O$. Our proposal is different from this and other TPM schemes, since we do not use any initial projective measurement and the initial state fully evolves according to its quantum dynamics. 
This is the typical situation encountered when considering the evolution of quantum systems, where the measurement is performed only at the final time {-- like during quantum computing algorithms.} Thus, analysing the differences and analogies between our scheme and other existing protocols helps in comparing the typical measurement procedures with those in quantum thermodynamics.

Remarkably, we are able to characterize the fluctuations of energy changes by distinguishing between contributions stemming from quantum coherences and those resulting from initial populations, albeit at the cost of a quantifiable extra uncertainty. These results offer the possibility to set coherence-induced quantum effects apart from those due to thermal fluctuations. Renouncing to the initial energy measurement on the system entails a substantive experimental simplification, thus making such approach an alternative to the TPM scheme when quantum signatures are considered. We demonstrate the effectiveness of EPM in pinpointing the role of initial coherences in the statistics of energy fluctuations by performing a series of experiments using the IBM Quantum Experience (IBMQ) platform. { This highlights the applicability of our scheme for the characterization of the energetics of quantum computation, a topic which is receiving growing attention in recent years~\cite{gardas2018quantum,PhysRevA.101.042106,buffoni2020thermodynamics,deffner2021energetic}.}

\noindent
{\bf Coherence in the energy eigenbasis.--} 
Let us consider a $d$-dimensional quantum system $\mathcal{S}$ evolving according to a one-parameter family of completely-positive and trace-preserving (CPTP) maps
$\Phi_t:\rho_{\rm i}\rightarrow\rho_{\rm f}=\Phi_t[\rho_{\rm i}]$~\cite{Caruso_RevModPhys_2014} within the time interval
$[t_{\rm i},t_{\rm f}]$. Here, $\rho_{\rm i}$ ($\rho_{\rm f}$) is the initial (final) density operator of the system. Our derivation can be specialized to the case of closed systems with time-dependent Hamiltonian, where energy fluctuations identify as work, or to open time-independent ones where only heat-transfer occurs. 

Let us thus consider a system ${\cal S}$ subject to \textit{no} initial projective measurement and characterize energy fluctuations only through a final-time measurement. 
The only energy measurement of our  protocol is performed at the final time $t_{\rm f}$. This generates the trajectories
${\cal T}^k_i:\,\rho_{\rm i}\rightarrow\Pi_{\rm f}^{k}$ with $\Pi_{\rm f}^{k} \equiv |E^{k}_{\rm f}\rangle\!\langle E^{k}_{\rm f}|$ denoting the projector onto the $k$-th energy eigenstates $\vert E^k_{\rm f}\rangle$ of the Hamiltonian at time $t_{\rm f}$, i.e., $H(t_{\rm f})=\sum_k
E_{\rm f}^k \Pi_{\rm f}^k$.
The stochasticity of the outcomes provided by the EPM protocol, with respect to the initial energies that ${\cal S}$ would have \textit{if} the energy had been measured, 
makes $\Delta E \equiv E_{\rm f} - E_{\rm i}$ a random variable. 

Dynamically, the initial quantum coherence in the state of ${\cal S}$, written in the energy basis, is accounted for by considering the probability distribution of the final energy due to the evolved initial state $\rho_{\rm i}$, comprising its coherence. By fixing the energy of $\mathcal{S}$ at $t_{\rm f}$, there is a probability law weighting the trajectories ${\cal T}^k_i$, which can be arranged in $N$ groups corresponding to the number of possible energy values at $t_{\rm i}$. This is a classical law, interpreted as the uncertainty on the values of $E_{\rm i}$, and thus $\Delta E$. By performing energy measurements at the final time $t_{\rm f}$, one can embed the effects of initial coherences into single realizations of the evolution. The uncertainty on $E_{\rm i}$ reflects the fact that its values are obtained as if we were performing a \textit{virtual} projective measurements, thus without any state collapse. This {entails} independence of the measurements at $t_\text{f}$ with respect to the initial virtual one.

Suppose the initial state $\rho_{\rm i}$ is not diagonal in the energy basis of ${\cal S}$: One can object that there is an observable $O$ on whose basis $\rho_{\rm i}$ is diagonal. However, there is an expected difference between the cases where \textit{a)} a measurement of $O$ is done at time $t_{\rm i}$\,, then one starts each trajectory from an eigenstate of $O$ and averages a posteriori over all possible results of the first measurement~\cite{MicadeiPRL2020}, and \textit{b)} no measurement is implemented and the dynamics can show interference in the energy basis. Such difference will be quantified later.

\begin{figure}[t]
\centering
\includegraphics[width=0.75\columnwidth]{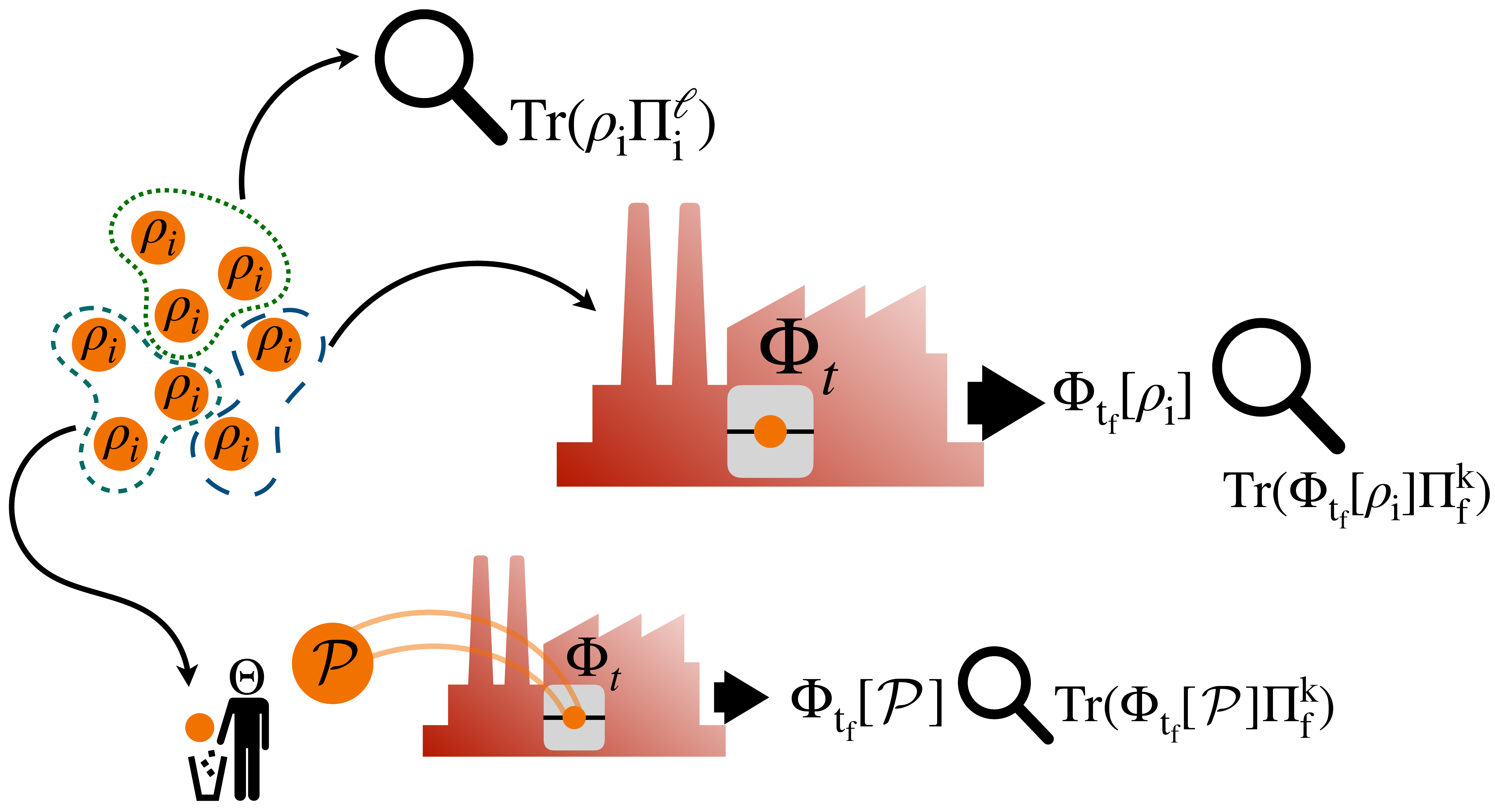}
\caption{Protocol for the quantification of energy fluctuations and the extraction of information about coherence. An ensemble of identical systems, prepared in the initial state $\rho_{\rm{i}}$, is divided in three subgroups. One is used to obtain $p^{\ell}_{\rm i}={\rm Tr}(\rho_i\Pi^{\ell}_{\rm i})$ via an initial energy measurement. The second goes through a dephasing channel, returning a state $\mathcal{P}$ diagonal in the energy basis. This then undergoes map $\Phi_t$ and is used to determine $p_{\mathcal{P}}^k={\rm Tr}(\Phi_{t_{\rm f}}[\mathcal{P}]\Pi^{k}_{\rm f})$. The systems in the third subgroup are not initially measured but  subjected to the dynamics and used to obtain $p_{\rm f}^{k}={\rm Tr}(\Phi_{t_{\rm f}}[\rho_i]\Pi^{k}_{\rm{f}})$.}
\label{fig:figprotocol}
\end{figure}

If the energy is not measured at $t_{\rm i}$, how can we talk about the initial energies $E_{\rm i}$? Such information, and the related thermodynamic cost, is encoded in $\rho_{\rm i}$, which is such that, \textit{if} we decide to measure the energy, we would find the initial energies $E_{\rm i}$. One could prepare $\rho_{\rm i}$ a large number of times, and in a fraction of them measure energy to verify that the eigenvalues $E_{\rm i}^{\ell}$'s of the Hamiltonian at $t=t_{\rm i}$ , i.e., $H(t_{\rm i})=\sum_\ell E_{\rm i}^\ell \Pi_{\rm i}^\ell$, are obtained with the probability assigned by $\rho_{\rm i}$ [cf. Fig.~\ref{fig:figprotocol}].
At the remaining times one uses $\rho_{\rm i}$ as input for our protocol \textit{without} measuring energy at $t_{\rm i}$.

\noindent
{\bf Energy-change distribution and link with fluctuation relations.--}
Let us assume a time-dependent Hamiltonian process and define the
probability distribution associated to $\Delta E$ by analyzing its properties. At the single-trajectory level, the density operator after the end-point energy measurement is one of the eigenstates $\Pi_{\rm f}^{k}$ of the time-dependent Hamiltonian $H(t_{\rm f})$. Such state is achieved with probability
\begin{equation}\label{p_f_k}
p_{\rm f}^{k} \equiv {\rm Tr}\left(\rho_{\rm f}\Pi_{\rm f}^{k}\right) = {\rm Tr}\left(\Phi_{t_{\rm f}}[\rho_{\rm i}]\Pi_{\rm f}^{k}\right).
\end{equation}
Thus, given the change
$\Delta E^{k,\ell} \equiv E^{k}_{\rm f} - E^{\ell}_{\rm i}$ in terms of the
eigenvalues of $H(t)$, the probability distribution of $\Delta E$
is 
\begin{equation}\label{prob_distr}
{\rm P}_{\rm coh}(\Delta E) = \sum_{k}p_{\rm f}^{k}\sum_{\ell}p_{\rm i}^{\ell}\delta(\Delta E - \Delta E^{k,\ell}),
\end{equation}
where $p_{\rm i}^{\ell} \equiv p(E_{\rm i}^{\ell}) = {\rm Tr}
(\rho_{\rm i}\Pi_{\rm i}^{\ell})$ is the probability of obtaining
$E^{\ell}_{\rm i}$ \textit{if} an energy measurement was performed
on $\mathcal{S}$ (initial virtual measurement).
In Eq.~\eqref{prob_distr}, the suffix "\rm{coh}" stands for "coherence".
The joint probability $p(E_{\rm i}^{\ell},E_{\rm f}^{k})$ associated to the
stochastic variable $\Delta E^{k,\ell}$, such that
${\rm P}_{\rm coh}(\Delta E) = \sum_{\ell,k} p(E_{\rm i}^{\ell},E_{\rm f}^{k})
\delta(\Delta E - \Delta E^{k,\ell})$,  
can then be written as
\begin{equation}\label{pdf}
  p(E_{\rm i}^{\ell},E_{\rm f}^{k})= p_{\rm i}^{\ell}p_{\rm f}^{k} =
  {\rm Tr}\left(\rho_{\rm i}\Pi_{\rm i}^{\ell}\right){\rm Tr}\left(\Phi_{t_{\rm f}}
  [\rho_{\rm i}]\Pi_{\rm f}^{k}\right) \equiv p_{\rm coh}^{\ell,k}.
\end{equation}
As already noticed, the assumption behind Eq.~\eqref{pdf} is the
statistical independence of the final energy projective
measurements and initial virtual one. This comes from the
fact that the initial measurement is not performed and only
the statistics related to the initial state preparation is used. The following properties hold:
 
\noindent
    {\bf Property (i)}\label{uno} ${\rm P}_{\rm coh}(\Delta E)$
    is 
    such that $\sum_{k,\ell}p_{\rm coh}^{\ell,k}=1$. \\
\noindent
    {\bf Property (ii)} The average energy variation
    $\langle\Delta E\rangle_{{\rm P}_{\rm coh}} \equiv \int d\Delta E~{\rm P}_{\rm coh}(\Delta E)\Delta E$ reproduces the average energy change induced by the CPTP map $\Phi_t$, that is
\begin{equation}\label{average_DeltaE}
  \langle\Delta E\rangle = {\rm Tr}(H(t_{\rm f})\rho_{\rm f}) - {\rm Tr}(H(t_{\rm i})\rho_{\rm i}),
\end{equation}
where we have assumed statistical independence between virtual initial energy measurements and final ones~\footnote{
  Let us observe that, in order to obtain Eq.~\eqref{average_DeltaE}, we need to weight the statistics of the measurement outcomes at $t=t_{\rm f}$ with the probabilities to initially get one of the outcomes $E_{\rm i}$. Otherwise the energy variation $\Delta E$ is erroneously proportional to
  ${\rm Tr}[H(t_{\rm f})\rho_{\rm f}]$.}. \\
\noindent
    {\bf Property (iii)} {$P_{{\rm coh}(\Delta E)}$ does not reduce to the TPM probability distribution for $[\rho_{\rm i},H(0)]=0$, i.e., it cannot result from a fluctuation theorem (FT) protocol in the sense of Ref.~\cite{LostaglioPRL2018}.}

Even by replacing the initial state $\rho_{\rm i}$ in
Eq.~\eqref{prob_distr} with a state diagonal in the (initial) energy basis, it is not possible to recover
the conventional energy-change statistics resulting from the TPM protocol. The latter is recovered only when the initial state is an energy eigenstate (cf. the Supplementary Material (SM) accompanying this paper~\cite{SM}).
{For an initial state diagonal in the energy eigenbasis}, the discrepancy between the {TPM and EPM} joint probabilities is due to classical uncertainty on the initial state of ${\cal S}$, which is retained in our scheme but is lost in TPM due to the initial energy measurement. As shown in Ref.~\cite{SM}, this agrees with the no-go theorem in Ref.~\cite{Perarnau-LlobetPRL2017}.
For the same reasons,
besides a few exceptions, the distribution
${\rm P}_{\rm coh}(\Delta E)$ may \emph{not} be convex
under a linear mixture of protocols that only differ
by the initial density operator $\rho_{\rm i}$~\cite{SM}. Therefore, given $\rho_{\rm i} = \zeta\rho_{\rm i, 1} + (1-\zeta)\rho_{\rm i, 2}$
with $\zeta\in[0,1]$, ${\rm P}_{\rm coh}(\Delta E|\rho_{\rm i})$
cannot in general be expressed as a
linear composition of the distributions
${\rm P}_{\rm coh}(\Delta E|\rho_{\rm i, 1})$
and ${\rm P}_{\rm coh}(\Delta E|\rho_{\rm i,2})$. 

In order to pinpoint the effect of coherence in the energy basis of $\rho_{\rm i}$ and separate it from classical uncertainty, we take $\rho_{\rm i}=\mathcal{P}+\chi$ with $\mathcal{P}$ diagonal in the energy basis and $\chi$ encoding the coherence contributions ($\rm Tr(\chi)=0$).
Then $p_{\rm coh}^{\ell,k}$ in Eq.~\eqref{pdf} can be split as $p_{\rm coh}^{\ell,k} = p_{\rm i}^{\ell}p_{\rm f}^{k} \equiv p^{\ell}_{\rm i}p_{\mathcal{P}}^k+p^{\ell}_{\rm i}p_{\chi}^{k}$ with
\begin{equation}
\label{splittato2}
p_{\rm f}^{k} \equiv p_{\mathcal{P}}^k+p_{\chi}^{k}={\rm Tr}(\Phi_{t_{\rm f}}[\mathcal{P}]\Pi^{k}_{\rm f})+{\rm Tr}(\Phi_{t_{\rm f}}[\chi]\Pi^{k}_{\rm f}).
\end{equation}

The term $p^{\ell}_{\rm i}p_{\mathcal{P}}^k$
encodes information on classical uncertainty on the initial system populations, while $p^{\ell}_{\rm i}p_{\chi}^{k}$
takes into account the effects of initial coherence. We introduce $p_{\rm coh}^{\mathcal{P}} \equiv p^{\ell}_{\rm i} p_{\mathcal{P}}^k$ and,
owing to the statistical independence of outcomes $\{E^{\ell}_{\rm i}\}$ and $\{E^{k}_{\rm f}\}$, such terms can be separately analyzed. In particular, the term containing information on the initial coherence can be determined as illustrated in
Fig.~\ref{fig:figprotocol}.

Note that the absence of initial coherences makes the EPM distribution equal to the product of the marginals of the
TPM distribution~\footnote{Similarly, the same result holds if we compare the probability density function of the EPM protocol, for a general initial state this time, with the one of the MLL scheme~\cite{MicadeiPRL2020} (see also the SM to this work). We thank Gabriel Landi for pointing out this result in relation to the MLL scheme.}. {We thus have $\mathcal{H}(p_{\rm{TPM}})\leq \mathcal{H}(p_{\rm{coh}}|_{\chi=0})$, where $\mathcal{H}(p)$ is the Shannon entropy of a generic distribution $p$. This inequality follows from the positivity of mutual information.} However, the same result is not true in general if initial coherence is present (cf. the case study of a three-level thermal engine in Ref.~\cite{SM}). 

We now address the differences with the protocol in Ref.~\cite{MicadeiPRL2020} -- which we label MLL -- to study the effects of coherence. In MLL, an  initial state decomposed in terms of its eigenstates $\{\vert s\rangle\}$ as $\rho_{\rm i}=\sum_s p^{s}|s\rangle\!\langle s|$,
is associated with the joint probability
$p_{\rm MLL}^{\ell,k}\equiv\sum_s p^{s} |\langle s|E^{\ell}_{\rm i}\rangle|^2{\rm Tr}(\Phi_{t_{\rm f}}[|s\rangle\!\langle s|]\Pi^{k}_{\rm{f}})$.
This reduces to the joint probability of the TPM protocol for $\rho_{\rm i}$ diagonal in the energy basis, and to the distribution $p_{\rm coh}^{\ell,k}$ of our protocol for initial pure states. However, for a generic initial state, such correspondences are lost and MLL requires $\rho_{\rm i}$ to be one of its eigenstates, as the construction of $p_{\rm{MLL}}^{\ell,k}$ requires to know the evolution of each component of $\rho_{\rm i}$. The EPM protocol thus requires less information on the dynamics at the cost of extra uncertainty on the statistics of $\Delta E$ (cf. Ref.~\cite{SM} for a comparison between EPM, MLL and TPM).

\noindent
{\bf Linear response approximation.--}
We now further characterize the distribution of energy changes and address its $1^\text{st}$
and $2^\text{nd}$ statistical moments. 
As with MLL, Eq.~\eqref{average_DeltaE} recovers
the expected difference of the averaged initial and final Hamiltonian. This is true in the TPM scheme only when the initial state is the mixture resulting from the first energy measurement. 
From Eq.~\eqref{prob_distr} one gets
\begin{equation}
\begin{aligned}
 \langle\Delta E^2\rangle &=\,\langle\Delta E^2\rangle_{\mathcal{P}}+{\rm Tr}({H}^2(t_{\rm f})\Phi_{t_{\rm f}}[\chi])\\ 
 &- 2\,{\rm Tr}(\Phi_{t_{\rm f}}[\chi]{H}(t_{\rm f}))\,{\rm Tr}(\mathcal{P}{H}(t_{\rm i})),
\end{aligned}\label{division}
\end{equation}
with $\langle\Delta E^2\rangle_{\mathcal{P}}$ given by assuming $\rho_{\rm i}\rightarrow \mathcal{P}$. Note that Eq.~\ref{division} coincides with the result of MLL (TPM) only if the initial state is pure (an eigenstate of ${H}(t_{\rm i})$). Moreover, if $\mathcal{P}$ is a projector, then
$\langle\Delta E^2\rangle_{\mathcal{P}}=\langle \Delta E^2\rangle_{\rm TPM}$
and all the differences in the $2^{\rm nd}$ moments are due to coherences in $\rho_{\rm i}$. The latter are unavoidably destroyed in the TPM protocol.

\noindent
{\bf Characteristic function and physical meaning.--}
The information about the statistics of the energy-change distribution is encoded in the characteristic function
$\mathcal{G}(u) \equiv \langle e^{iu\Delta E}\rangle_{{\rm P}_{\rm coh}} = \int d\Delta E\,e^{iu\Delta E}{\rm P}_{\rm coh}(\Delta E)$ corresponding to the distribution
${\rm P}_{\rm coh}(\Delta E)$. As the outcomes $\{E_{\rm f}^{(k)}\}$
of the final energy  measurement are statistically independent
from the initial virtual ones $\{E_{\rm i}^{(\ell)}\}$, we have
\begin{equation}\label{G_u}
    \mathcal{G}(u) = {\rm Tr}(e^{-iu{H}(t_{\rm i})}\rho_{\rm i})\,{\rm Tr}(e^{iu{H}(t_{\rm f})}\Phi_{t_{\rm f}}[\rho_{\rm i}]),
\end{equation}
showing that the fluctuations of $\Delta E$ originate
both from the action of map $\Phi_t[\rho]$ on the initial state of ${\cal S}$ and the uncertainty in its energy at $t=t_{\rm i}$.
\begin{figure}[t!]

\centering
\includegraphics[width=0.95\columnwidth]{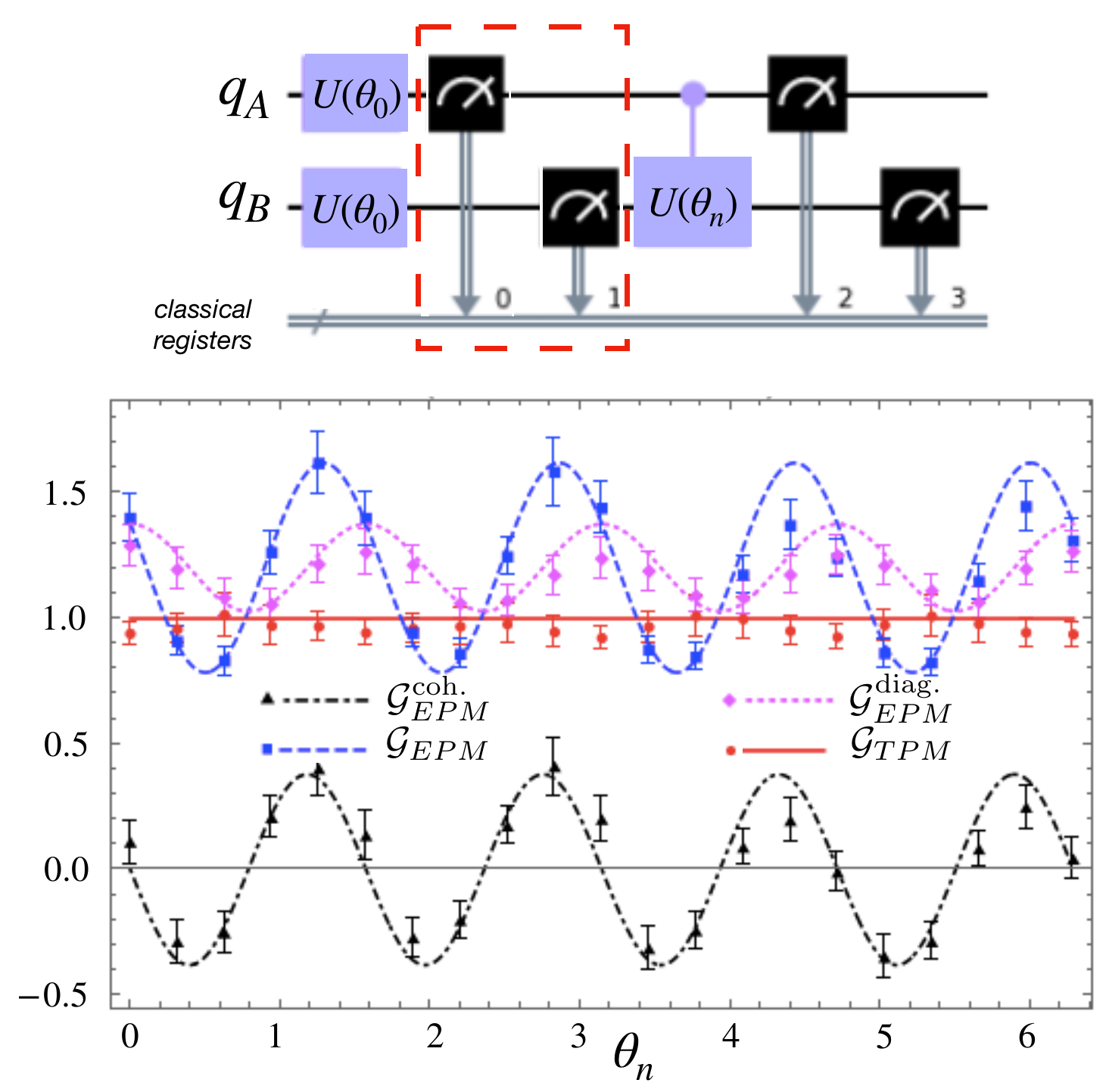}
\caption{\textbf{Top:} \textit{Circuits implemented in IBMQ}. The initial state 
is prepared by applying two identical single-qubit gates $U(\theta_{0})$ onto $|00\rangle$ (we use $\theta_0=2$~\cite{SM}). In TPM, two initial projective measurements destroy any coherence in the computational basis, while in EPM such measurements (enclosed in the dashed red box) are absent. We then implement the controlled gate $U(\theta_{n})$, with $\theta_n \equiv n\pi/10$ and $n=0,\ldots,20$, followed by two projective measurements in the computational basis. The results are stored in four classical registers to allow the analysis of the energy change statistics. \textbf{Bottom:} \textit{Comparison of the characteristic functions for EPM and TPM}. The lines show the theoretical predictions, while the points (with their error bars) the experimental results. Each data point has been obtained from 2048 experimental runs. The solid red line and circles are related to the results obtained by applying  TPM. The dashed blue line and squared refer to the EPM characteristic function. Finally, the dotted magenta line and rhombuses (dot-dashed black line and triangles) show the contribution of the diagonal (off-diagonal) parts of the initial state $\rho_{\rm i}$ in the computational basis. The inverse (physical) temperature of the diagonal part of the initial state is $\beta=0.443/\epsilon$ where $\epsilon\sim 5$~MHz is the energy gap for the superconducting qubits, as provided by the IBMQ documentation. 
}
\label{fig:circuit}
\end{figure}
We now highlight the deviation of the EPM-inferred statistics
from a standard FT~\cite{CampisiRMP2011,EspositoRMP2009}.
We consider $\mathcal{G}(i\beta)$,
where $\beta$ is a \textit{reference} inverse temperature
(taken as a free parameter), and introduce the 
reference equilibrium states
$\rho_{\rm i(f)}^{\rm th} \equiv e^{-\beta {H}({ t}_{\rm i(f)})}/Z_{\rm i(f)}$ with
$Z_{\rm i(f)} \equiv {\rm Tr}(e^{-\beta {H}({\rm t}_{\rm i(f)})})$. For $\rho_{\rm i} = \rho_{\rm i}^{\rm th}+\chi$ we get
\begin{equation}\label{eq:Jarzynski_like_rel}
    \langle e^{-\beta(\Delta E{-}\Delta F)}\rangle{=}d\left[{\rm Tr}\left(\rho_{\rm f}^{\rm th}\,\Phi_{t_{\rm f}}[\rho_{\rm i}^{\rm th}]\right){+}{\rm Tr}\left(\rho_{\rm f}^{\rm th}\,\Phi_{t_{\rm f}}[\chi]\right)\right],
\end{equation}
with $\Delta F$ the free energy difference {and $d$ the dimension of the Hilbert space of ${\cal S}$} (cf.\,Ref.~\cite{SM} for details). Eq.~\eqref{eq:Jarzynski_like_rel} deviates from unity, {i.e.} from a standard fluctuation theorem, even for unital channels and due to two terms. The first,
$d\,{\rm Tr}(\rho_{\rm f}^{\rm th}\,\Phi_{t_{\rm f}}[\rho_{\rm i}^{\rm th}])$, is the additional uncertainty introduced by not performing the initial energy measurement and is present even for $\chi=0$. The second 
quantifies the deviation due to initial quantum coherences and bridges stochastic thermodynamics and quantum signatures of open dynamics. Eq.~\eqref{eq:Jarzynski_like_rel} is thus one of the main results of this paper.

\noindent
{\bf Experimental Results.--}
To illustrate experimentally the power and versatility of EPM,
we make use of the IBMQ platform. In particular, we perform a series of experiments based on the use of a two-qubit gate, by following the protocol illustrated in Fig.~\ref{fig:figprotocol} 
for the extraction of initial coherence contributions.

On the IBMQ quantum computer, we implement a two-qubit circuit with initial (pure) separable state $\rho_{\rm i} = \rho_{\rm i}^{\rm th}+\chi$, where $\rho_{\rm i}^{\rm th}=e^{-\beta (H_A+H_B)}/Z$ (with $Z=\text{tr}[\rho_{\rm i}^{\rm th}]$ and inverse temperature $\beta$) is a thermal state of the local Hamiltonian $H_A+H_B=\epsilon(\sigma_z^{(A)}+\sigma_z^{(B)})$ of the two qubits ($\epsilon\sim5$MHz is the energy gap between the logical states of each superconducting qubit). Here, $\rho_{\rm i}^{\rm th}$ is  diagonal in the computational basis, while $\chi$  stands for the initial coherence in this basis. Such initial state can be easily prepared starting from the default configuration of the logical qubit of the IBMQ device by way of properly designed single-qubit gates (cf. Fig.~\ref{fig:circuit} and Ref.~\cite{SM}). 

The top panel of Fig.~\ref{fig:circuit} shows the circuit implemented in the IBMQ. After the initialization, the circuit performs a controlled gate. The difference between EPM and TPM is in the absence of the first two projective measurements (red box in the figure) for the former. Then, we repeat the experiments by varying one of the parameters of the controlled gate.
It is worth noticing that, 
{ while an ``effective'' Hamiltonian of the circuit could be obtained by reverse engineering the implemented unitary evolution, the IBMQ does not enable to directly measure it, as only local measurements of $\sigma_{z}$ (and, thus, of the qubits local energies) are allowed. 
Thus, in analogy with the experiment in Ref.~\cite{cimini2020experimental}, just the statistics of the local energy fluctuations are taken into account.}

In the bottom panel of Fig.~\ref{fig:circuit}, we consider the deviation of $\langle e^{-\beta(\Delta E-\Delta F)}\rangle$ from unity when using the EPM protocol. In the considered case, the free energy variation vanishes. Thus, we are comparing the characteristic functions, evaluated at $u=i\beta$, of EPM and TPM. The Jarzynski identity $\mathcal{G}_{\rm TPM}(i\beta)=1$, stemming from TPM, is nicely recovered from the experimental data. This is compared to the contributions in Eq.~\eqref{eq:Jarzynski_like_rel} linked to the diagonal and off-diagonal parts of the initial state. For the case investigated here, we observe a non-negligible contribution from the initial coherence $\chi$ of $\rho_{\rm i}$, and a clear discrepancy between the TPM result and the contribution to the EPM characteristic function depending on the (thermal) diagonal part $\rho_{\rm i}^{\rm th}$ of $\rho_{\rm i}$. {As stressed above, such a discrepancy originates from the additional uncertainty on the initial energies introduced by our protocol.} Moreover, the statistics of energy changes in Fig.~\ref{fig:circuit} can be reproduced to a good approximation by looking at just the firsts two moments of the EPM (or TPM) distribution~\cite{SM}. Therefore, an analysis in linear approximation is able to capture the main features of the energy fluctuations that pertain to the quantum circuits under scrutiny.

\noindent
{\bf Conclusions.--}
We have introduced an EPM protocol for the evaluation of the energy-change fluctuations that takes into account the presence of quantum coherence in the initial state of the system. The protocol does not require information on the dynamics nor special preparations, which casts it apart from other schemes~\cite{DeffnerPRE2016,MicadeiPRL2020,Sone2020}, and solely relies on the final energy measurement. 
The EPM approach could be more conducive of experimental validation than the notoriously challenging TPM one, and could thus enlarge the range of systems whose energy fluctuations could be tested.
{For instance, quantum computing platforms present a natural arena in which the methods developed in this work could find fruitful applications, as showcased by our analysis of the IBMQ two-qubit logic circuit. Indeed, the EPM approach not only allows to account for the effect of the initial coherence but also resemble the way in which quantum computing algorithms are actually performed, where only a final measurement is present. Furthermore, the EPM approach may also come in handy for systems with degenerate energy levels, as in many-body physics. Indeed, for initial states involving only levels within degenerate subspaces and a dynamics that leaves the latter invariant, the TPM scheme would return vanishing energy fluctuations. In contrast, our EPM would allow for the characterization of the energy change statistics resulting from the initial coherence alone.}%

\paragraph*{Acknowledgments.--}The authors gratefully acknowledge L. Buffoni, N. Fabbri, S. Hern\'andez-G\'omez, G.T. Landi, M. Lostaglio, S. Martina and F. Poggiali for fruitful discussions and comments. This work was supported by MISTI Global Seed Funds MIT-FVG Collaboration Grant "NV centers for the test of the Quantum Jarzynski Equality (NVQJE)", H2020-FETOPEN-2018-2020 project PATHOS (grant nr.\,828946), UNIFI grant Q-CODYCES, the MSCA IF project pERFEcTO (grant nr. 795782), the DeutscheForschungsgemeinschaft (DFG, German Research Founda-tion) project number BR 5221/4-1, the H2020-FETOPEN-2018-2020 project TEQ (grant nr.~766900), the DfE-SFI Investigator Programme (grant 15/IA/2864), COST Action CA15220, the Royal Society Wolfson Research Fellowship (RSWF\textbackslash R3\textbackslash183013), the Royal Society International Exchanges Programme (IEC\textbackslash R2\textbackslash192220), the Leverhulme Trust Research Project Grant (grant nr.~RGP-2018-266) and the CNR/RS (London) project "Testing fundamental theories with ultracold atoms". We acknowledge the use of IBM Quantum services for this work. The views expressed are those of the authors, and do not reflect the official policy or position of IBM or the IBM Quantum team.

\bibliography{biblio_AT}

\begin{thebibliography}{40}%
\makeatletter
\providecommand \@ifxundefined [1]{%
 \@ifx{#1\undefined}
}%
\providecommand \@ifnum [1]{%
 \ifnum #1\expandafter \@firstoftwo
 \else \expandafter \@secondoftwo
 \fi
}%
\providecommand \@ifx [1]{%
 \ifx #1\expandafter \@firstoftwo
 \else \expandafter \@secondoftwo
 \fi
}%
\providecommand \natexlab [1]{#1}%
\providecommand \enquote  [1]{``#1''}%
\providecommand \bibnamefont  [1]{#1}%
\providecommand \bibfnamefont [1]{#1}%
\providecommand \citenamefont [1]{#1}%
\providecommand \href@noop [0]{\@secondoftwo}%
\providecommand \href [0]{\begingroup \@sanitize@url \@href}%
\providecommand \@href[1]{\@@startlink{#1}\@@href}%
\providecommand \@@href[1]{\endgroup#1\@@endlink}%
\providecommand \@sanitize@url [0]{\catcode `\\12\catcode `\$12\catcode
  `\&12\catcode `\#12\catcode `\^12\catcode `\_12\catcode `\%12\relax}%
\providecommand \@@startlink[1]{}%
\providecommand \@@endlink[0]{}%
\providecommand \url  [0]{\begingroup\@sanitize@url \@url }%
\providecommand \@url [1]{\endgroup\@href {#1}{\urlprefix }}%
\providecommand \urlprefix  [0]{URL }%
\providecommand \Eprint [0]{\href }%
\providecommand \doibase [0]{http://dx.doi.org/}%
\providecommand \selectlanguage [0]{\@gobble}%
\providecommand \bibinfo  [0]{\@secondoftwo}%
\providecommand \bibfield  [0]{\@secondoftwo}%
\providecommand \translation [1]{[#1]}%
\providecommand \BibitemOpen [0]{}%
\providecommand \bibitemStop [0]{}%
\providecommand \bibitemNoStop [0]{.\EOS\space}%
\providecommand \EOS [0]{\spacefactor3000\relax}%
\providecommand \BibitemShut  [1]{\csname bibitem#1\endcsname}%
\let\auto@bib@innerbib\@empty
\bibitem [{\citenamefont {Vinjanampathy}\ and\ \citenamefont
  {Anders}(2016)}]{Vinjanampathy2016}%
  \BibitemOpen
  \bibfield  {author} {\bibinfo {author} {\bibfnamefont {S.}~\bibnamefont
  {Vinjanampathy}}\ and\ \bibinfo {author} {\bibfnamefont {J.}~\bibnamefont
  {Anders}},\ }\href {https://doi.org/10.1080/00107514.2016.1201896} {\bibfield
   {journal} {\bibinfo  {journal} {Contemp. Phys. {\bf 57}, 545}\ } (\bibinfo
  {year} {2016})}\BibitemShut {NoStop}%
\bibitem [{\citenamefont {Sagawa}(2013)}]{Sagawa2014}%
  \BibitemOpen
  \bibfield  {author} {\bibinfo {author} {\bibfnamefont {T.}~\bibnamefont
  {Sagawa}},\ }in\ \href@noop {} {\emph {\bibinfo {booktitle} {Lectures on
  Quantum Computing, Thermodynamics and Statistical Physics}}},\ \bibinfo
  {editor} {edited by\ \bibinfo {editor} {\bibfnamefont {M.}~\bibnamefont
  {Nakahara}}\ and\ \bibinfo {editor} {\bibfnamefont {S.}~\bibnamefont
  {Tanaka}}}\ (\bibinfo  {publisher} {World Scientific Publishing Co. Pte.
  Ltd.},\ \bibinfo {year} {2013})\BibitemShut {NoStop}%
\bibitem [{Boo(2019)}]{BookQT}%
  \BibitemOpen
  in\ \href@noop {} {\emph {\bibinfo {booktitle} {Thermodynamics in the Quantum
  Regime}}},\ \bibinfo {editor} {edited by\ \bibinfo {editor} {\bibfnamefont
  {F.}~\bibnamefont {Binder}}, \bibinfo {editor} {\bibfnamefont
  {L.}~\bibnamefont {Correa}}, \bibinfo {editor} {\bibfnamefont
  {C.}~\bibnamefont {Gogolin}}, \bibinfo {editor} {\bibfnamefont
  {J.}~\bibnamefont {Anders}}, \ and\ \bibinfo {editor} {\bibfnamefont
  {G.}~\bibnamefont {Adesso}}}\ (\bibinfo  {publisher} {Springer International
  Publishing},\ \bibinfo {year} {2019})\BibitemShut {NoStop}%
\bibitem [{\citenamefont {Deffner}\ and\ \citenamefont
  {Campbell}(2019)}]{deffner2019qtd}%
  \BibitemOpen
  \bibfield  {author} {\bibinfo {author} {\bibfnamefont {S.}~\bibnamefont
  {Deffner}}\ and\ \bibinfo {author} {\bibfnamefont {S.}~\bibnamefont
  {Campbell}},\ }\href@noop {} {\emph {\bibinfo {title} {Quantum
  Thermodynamics: An introduction to the thermodynamics of quantum
  information}}}\ (\bibinfo  {publisher} {Morgan \& Claypool Publishers},\
  \bibinfo {year} {2019})\BibitemShut {NoStop}%
\bibitem [{\citenamefont {Talkner}\ \emph {et~al.}(2007)\citenamefont
  {Talkner}, \citenamefont {Lutz},\ and\ \citenamefont
  {H\"anggi}}]{TalknerPRE2007}%
  \BibitemOpen
  \bibfield  {author} {\bibinfo {author} {\bibfnamefont {P.}~\bibnamefont
  {Talkner}}, \bibinfo {author} {\bibfnamefont {E.}~\bibnamefont {Lutz}}, \
  and\ \bibinfo {author} {\bibfnamefont {P.}~\bibnamefont {H\"anggi}},\ }\href
  {\doibase 10.1103/PhysRevE.75.050102} {\bibfield  {journal} {\bibinfo
  {journal} {Phys. Rev. E}\ }\textbf {\bibinfo {volume} {75}},\ \bibinfo
  {pages} {050102} (\bibinfo {year} {2007})}\BibitemShut {NoStop}%
\bibitem [{\citenamefont {Esposito}\ \emph {et~al.}(2009)\citenamefont
  {Esposito}, \citenamefont {Harbola},\ and\ \citenamefont
  {Mukamel}}]{EspositoRMP2009}%
  \BibitemOpen
  \bibfield  {author} {\bibinfo {author} {\bibfnamefont {M.}~\bibnamefont
  {Esposito}}, \bibinfo {author} {\bibfnamefont {U.}~\bibnamefont {Harbola}}, \
  and\ \bibinfo {author} {\bibfnamefont {S.}~\bibnamefont {Mukamel}},\ }\href
  {https://link.aps.org/doi/10.1103/RevModPhys.81.1665} {\bibfield  {journal}
  {\bibinfo  {journal} {Rev. Mod. Phys.}\ }\textbf {\bibinfo {volume} {81}},\
  \bibinfo {pages} {1665} (\bibinfo {year} {2009})}\BibitemShut {NoStop}%
\bibitem [{\citenamefont {Campisi}\ \emph {et~al.}(2011)\citenamefont
  {Campisi}, \citenamefont {H\"{a}nggi},\ and\ \citenamefont
  {Talkner}}]{CampisiRMP2011}%
  \BibitemOpen
  \bibfield  {author} {\bibinfo {author} {\bibfnamefont {M.}~\bibnamefont
  {Campisi}}, \bibinfo {author} {\bibfnamefont {P.}~\bibnamefont {H\"{a}nggi}},
  \ and\ \bibinfo {author} {\bibfnamefont {P.}~\bibnamefont {Talkner}},\ }\href
  {https://link.aps.org/doi/10.1103/RevModPhys.83.771} {\bibfield  {journal}
  {\bibinfo  {journal} {Rev. Mod. Phys.}\ }\textbf {\bibinfo {volume} {83}},\
  \bibinfo {pages} {771} (\bibinfo {year} {2011})}\BibitemShut {NoStop}%
\bibitem [{\citenamefont {Jacobs}(2014)}]{BookJacobs}%
  \BibitemOpen
  \bibfield  {author} {\bibinfo {author} {\bibfnamefont {K.}~\bibnamefont
  {Jacobs}},\ }\href@noop {} {\emph {\bibinfo {title} {Quantum Measurement
  Theory and its Applications}}}\ (\bibinfo  {publisher} {Cambridge University
  Press},\ \bibinfo {year} {2014})\BibitemShut {NoStop}%
\bibitem [{\citenamefont {Allahverdyan}(2014)}]{AllahverdyanPRE2014}%
  \BibitemOpen
  \bibfield  {author} {\bibinfo {author} {\bibfnamefont {A.}~\bibnamefont
  {Allahverdyan}},\ }\href
  {https://link.aps.org/doi/10.1103/PhysRevE.90.032137} {\bibfield  {journal}
  {\bibinfo  {journal} {Phys. Rev. E}\ }\textbf {\bibinfo {volume} {90}},\
  \bibinfo {pages} {032137} (\bibinfo {year} {2014})}\BibitemShut {NoStop}%
\bibitem [{\citenamefont {Lostaglio}\ \emph {et~al.}(2015)\citenamefont
  {Lostaglio}, \citenamefont {Jennings},\ and\ \citenamefont
  {Rudolph}}]{LostaglioNATCOMM2015}%
  \BibitemOpen
  \bibfield  {author} {\bibinfo {author} {\bibfnamefont {M.}~\bibnamefont
  {Lostaglio}}, \bibinfo {author} {\bibfnamefont {D.}~\bibnamefont {Jennings}},
  \ and\ \bibinfo {author} {\bibfnamefont {T.}~\bibnamefont {Rudolph}},\ }\href
  {https://doi.org/10.1038/ncomms7383} {\bibfield  {journal} {\bibinfo
  {journal} {Nat. Commun.}\ }\textbf {\bibinfo {volume} {6}},\ \bibinfo {pages}
  {6383} (\bibinfo {year} {2015})}\BibitemShut {NoStop}%
\bibitem [{\citenamefont {Mazzola}\ \emph {et~al.}(2013)\citenamefont
  {Mazzola}, \citenamefont {De~Chiara},\ and\ \citenamefont
  {Paternostro}}]{Mazzola2013}%
  \BibitemOpen
  \bibfield  {author} {\bibinfo {author} {\bibfnamefont {L.}~\bibnamefont
  {Mazzola}}, \bibinfo {author} {\bibfnamefont {G.}~\bibnamefont {De~Chiara}},
  \ and\ \bibinfo {author} {\bibfnamefont {M.}~\bibnamefont {Paternostro}},\
  }\href {https://link.aps.org/doi/10.1103/PhysRevLett.110.230602} {\bibfield
  {journal} {\bibinfo  {journal} {Phys. Rev. Lett.}\ }\textbf {\bibinfo
  {volume} {110}},\ \bibinfo {pages} {230602} (\bibinfo {year}
  {2013})}\BibitemShut {NoStop}%
\bibitem [{\citenamefont {Dorner}\ \emph {et~al.}(2013)\citenamefont {Dorner},
  \citenamefont {Clark}, \citenamefont {Heaney}, \citenamefont {Fazio},
  \citenamefont {Goold},\ and\ \citenamefont {Vedral}}]{Dorner2013}%
  \BibitemOpen
  \bibfield  {author} {\bibinfo {author} {\bibfnamefont {R.}~\bibnamefont
  {Dorner}}, \bibinfo {author} {\bibfnamefont {S.~R.}\ \bibnamefont {Clark}},
  \bibinfo {author} {\bibfnamefont {L.}~\bibnamefont {Heaney}}, \bibinfo
  {author} {\bibfnamefont {R.}~\bibnamefont {Fazio}}, \bibinfo {author}
  {\bibfnamefont {J.}~\bibnamefont {Goold}}, \ and\ \bibinfo {author}
  {\bibfnamefont {V.}~\bibnamefont {Vedral}},\ }\href
  {https://link.aps.org/doi/10.1103/PhysRevLett.110.230601} {\bibfield
  {journal} {\bibinfo  {journal} {Phys. Rev. Lett.}\ }\textbf {\bibinfo
  {volume} {110}},\ \bibinfo {pages} {230601} (\bibinfo {year}
  {2013})}\BibitemShut {NoStop}%
\bibitem [{\citenamefont {Batalh{\~a}o}\ \emph {et~al.}(2014)\citenamefont
  {Batalh{\~a}o}, \citenamefont {Souza}, \citenamefont {Mazzola}, \citenamefont
  {Auccaise}, \citenamefont {Sarthour}, \citenamefont {Oliveira}, \citenamefont
  {Goold}, \citenamefont {De~Chiara}, \citenamefont {Paternostro},\ and\
  \citenamefont {Serra}}]{Batalhao2014}%
  \BibitemOpen
  \bibfield  {author} {\bibinfo {author} {\bibfnamefont {T.~B.}\ \bibnamefont
  {Batalh{\~a}o}}, \bibinfo {author} {\bibfnamefont {A.~M.}\ \bibnamefont
  {Souza}}, \bibinfo {author} {\bibfnamefont {L.}~\bibnamefont {Mazzola}},
  \bibinfo {author} {\bibfnamefont {R.}~\bibnamefont {Auccaise}}, \bibinfo
  {author} {\bibfnamefont {R.~S.}\ \bibnamefont {Sarthour}}, \bibinfo {author}
  {\bibfnamefont {I.~S.}\ \bibnamefont {Oliveira}}, \bibinfo {author}
  {\bibfnamefont {J.}~\bibnamefont {Goold}}, \bibinfo {author} {\bibfnamefont
  {G.}~\bibnamefont {De~Chiara}}, \bibinfo {author} {\bibfnamefont
  {M.}~\bibnamefont {Paternostro}}, \ and\ \bibinfo {author} {\bibfnamefont
  {R.~M.}\ \bibnamefont {Serra}},\ }\href
  {https://link.aps.org/doi/10.1103/PhysRevLett.113.140601} {\bibfield
  {journal} {\bibinfo  {journal} {Phys. Rev. Lett.}\ }\textbf {\bibinfo
  {volume} {113}},\ \bibinfo {pages} {140601} (\bibinfo {year}
  {2014})}\BibitemShut {NoStop}%
\bibitem [{\citenamefont {Solinas}\ and\ \citenamefont
  {Gasparinetti}(2015)}]{SolinasPRE2015}%
  \BibitemOpen
  \bibfield  {author} {\bibinfo {author} {\bibfnamefont {P.}~\bibnamefont
  {Solinas}}\ and\ \bibinfo {author} {\bibfnamefont {S.}~\bibnamefont
  {Gasparinetti}},\ }\href
  {https://link.aps.org/doi/10.1103/PhysRevE.92.042150} {\bibfield  {journal}
  {\bibinfo  {journal} {Phys. Rev. E}\ }\textbf {\bibinfo {volume} {92}},\
  \bibinfo {pages} {042150} (\bibinfo {year} {2015})}\BibitemShut {NoStop}%
\bibitem [{\citenamefont {Solinas}\ and\ \citenamefont
  {Gasparinetti}(2016)}]{SolinasPRA2016}%
  \BibitemOpen
  \bibfield  {author} {\bibinfo {author} {\bibfnamefont {P.}~\bibnamefont
  {Solinas}}\ and\ \bibinfo {author} {\bibfnamefont {S.}~\bibnamefont
  {Gasparinetti}},\ }\href
  {https://link.aps.org/doi/10.1103/PhysRevA.94.052103} {\bibfield  {journal}
  {\bibinfo  {journal} {Phys. Rev. A}\ }\textbf {\bibinfo {volume} {94}},\
  \bibinfo {pages} {052103} (\bibinfo {year} {2016})}\BibitemShut {NoStop}%
\bibitem [{\citenamefont {Alhambra}\ \emph {et~al.}(2016)\citenamefont
  {Alhambra}, \citenamefont {Masanes}, \citenamefont {Oppenheim},\ and\
  \citenamefont {Perry}}]{AlhambraPRX2016}%
  \BibitemOpen
  \bibfield  {author} {\bibinfo {author} {\bibfnamefont {A.}~\bibnamefont
  {Alhambra}}, \bibinfo {author} {\bibfnamefont {L.}~\bibnamefont {Masanes}},
  \bibinfo {author} {\bibfnamefont {J.}~\bibnamefont {Oppenheim}}, \ and\
  \bibinfo {author} {\bibfnamefont {C.}~\bibnamefont {Perry}},\ }\href
  {https://link.aps.org/doi/10.1103/PhysRevX.6.041017} {\bibfield  {journal}
  {\bibinfo  {journal} {Phys. Rev. X}\ }\textbf {\bibinfo {volume} {6}},\
  \bibinfo {pages} {041017} (\bibinfo {year} {2016})}\BibitemShut {NoStop}%
\bibitem [{\citenamefont {\AA{}berg}(2018)}]{AbergPRX2018}%
  \BibitemOpen
  \bibfield  {author} {\bibinfo {author} {\bibfnamefont {J.}~\bibnamefont
  {\AA{}berg}},\ }\href {https://link.aps.org/doi/10.1103/PhysRevX.8.011019}
  {\bibfield  {journal} {\bibinfo  {journal} {Phys. Rev. X}\ }\textbf {\bibinfo
  {volume} {8}},\ \bibinfo {pages} {011019} (\bibinfo {year}
  {2018})}\BibitemShut {NoStop}%
\bibitem [{\citenamefont {Lostaglio}(2018)}]{LostaglioPRL2018}%
  \BibitemOpen
  \bibfield  {author} {\bibinfo {author} {\bibfnamefont {M.}~\bibnamefont
  {Lostaglio}},\ }\href
  {https://link.aps.org/doi/10.1103/PhysRevLett.120.040602} {\bibfield
  {journal} {\bibinfo  {journal} {Phys. Rev. Lett.}\ }\textbf {\bibinfo
  {volume} {120}},\ \bibinfo {pages} {040602} (\bibinfo {year}
  {2018})}\BibitemShut {NoStop}%
\bibitem [{\citenamefont {Xu}\ \emph {et~al.}(2018)\citenamefont {Xu},
  \citenamefont {Zou}, \citenamefont {Guo},\ and\ \citenamefont
  {Kong}}]{XuPRA2018}%
  \BibitemOpen
  \bibfield  {author} {\bibinfo {author} {\bibfnamefont {B.-M.}\ \bibnamefont
  {Xu}}, \bibinfo {author} {\bibfnamefont {J.}~\bibnamefont {Zou}}, \bibinfo
  {author} {\bibfnamefont {L.-S.}\ \bibnamefont {Guo}}, \ and\ \bibinfo
  {author} {\bibfnamefont {X.-M.}\ \bibnamefont {Kong}},\ }\href
  {https://link.aps.org/doi/10.1103/PhysRevA.97.052122} {\bibfield  {journal}
  {\bibinfo  {journal} {Phys. Rev. A}\ }\textbf {\bibinfo {volume} {97}},\
  \bibinfo {pages} {052122} (\bibinfo {year} {2018})}\BibitemShut {NoStop}%
\bibitem [{\citenamefont {Francica}\ \emph {et~al.}(2019)\citenamefont
  {Francica}, \citenamefont {Goold},\ and\ \citenamefont
  {Plastina}}]{FrancicaPRE2019}%
  \BibitemOpen
  \bibfield  {author} {\bibinfo {author} {\bibfnamefont {G.}~\bibnamefont
  {Francica}}, \bibinfo {author} {\bibfnamefont {J.}~\bibnamefont {Goold}}, \
  and\ \bibinfo {author} {\bibnamefont {Plastina}},\ }\href
  {https://link.aps.org/doi/10.1103/PhysRevE.99.042105} {\bibfield  {journal}
  {\bibinfo  {journal} {Phys. Rev. E}\ }\textbf {\bibinfo {volume} {99}},\
  \bibinfo {pages} {042105} (\bibinfo {year} {2019})}\BibitemShut {NoStop}%
\bibitem [{\citenamefont {Santos}\ \emph {et~al.}(2019)\citenamefont {Santos},
  \citenamefont {Celeri}, \citenamefont {Landi},\ and\ \citenamefont
  {Paternostro}}]{SantosnpjQI2019}%
  \BibitemOpen
  \bibfield  {author} {\bibinfo {author} {\bibfnamefont {J.}~\bibnamefont
  {Santos}}, \bibinfo {author} {\bibfnamefont {L.}~\bibnamefont {Celeri}},
  \bibinfo {author} {\bibfnamefont {G.}~\bibnamefont {Landi}}, \ and\ \bibinfo
  {author} {\bibfnamefont {M.}~\bibnamefont {Paternostro}},\ }\href
  {https://www.nature.com/articles/s41534-019-0138-y} {\bibfield  {journal}
  {\bibinfo  {journal} {npj Quant. Inf.}\ }\textbf {\bibinfo {volume} {5}},\
  \bibinfo {pages} {23} (\bibinfo {year} {2019})}\BibitemShut {NoStop}%
\bibitem [{\citenamefont {Mingo}\ and\ \citenamefont
  {Jennings}(2019)}]{MingoQuantum2019}%
  \BibitemOpen
  \bibfield  {author} {\bibinfo {author} {\bibfnamefont {E.~H.}\ \bibnamefont
  {Mingo}}\ and\ \bibinfo {author} {\bibfnamefont {D.}~\bibnamefont
  {Jennings}},\ }\href {https://doi.org/10.22331/q-2019-11-11-202} {\bibfield
  {journal} {\bibinfo  {journal} {Quantum}\ }\textbf {\bibinfo {volume} {3}},\
  \bibinfo {pages} {202} (\bibinfo {year} {2019})}\BibitemShut {NoStop}%
\bibitem [{\citenamefont {Kwon}\ and\ \citenamefont
  {Kim}(2019)}]{PhysRevX.9.031029}%
  \BibitemOpen
  \bibfield  {author} {\bibinfo {author} {\bibfnamefont {H.}~\bibnamefont
  {Kwon}}\ and\ \bibinfo {author} {\bibfnamefont {M.}~\bibnamefont {Kim}},\
  }\href {\doibase 10.1103/PhysRevX.9.031029} {\bibfield  {journal} {\bibinfo
  {journal} {Phys. Rev. X}\ }\textbf {\bibinfo {volume} {9}},\ \bibinfo {pages}
  {031029} (\bibinfo {year} {2019})}\BibitemShut {NoStop}%
\bibitem [{\citenamefont {Micadei}\ \emph {et~al.}(2020)\citenamefont
  {Micadei}, \citenamefont {Landi},\ and\ \citenamefont
  {Lutz}}]{MicadeiPRL2020}%
  \BibitemOpen
  \bibfield  {author} {\bibinfo {author} {\bibfnamefont {K.}~\bibnamefont
  {Micadei}}, \bibinfo {author} {\bibfnamefont {G.~T.}\ \bibnamefont {Landi}},
  \ and\ \bibinfo {author} {\bibfnamefont {E.}~\bibnamefont {Lutz}},\ }\href
  {https://link.aps.org/doi/10.1103/PhysRevLett.124.090602} {\bibfield
  {journal} {\bibinfo  {journal} {Phys. Rev. Lett.}\ }\textbf {\bibinfo
  {volume} {124}},\ \bibinfo {pages} {090602} (\bibinfo {year}
  {2020})}\BibitemShut {NoStop}%
\bibitem [{\citenamefont {Levy}\ and\ \citenamefont
  {Lostaglio}(2020)}]{LevyArxiv2019}%
  \BibitemOpen
  \bibfield  {author} {\bibinfo {author} {\bibfnamefont {A.}~\bibnamefont
  {Levy}}\ and\ \bibinfo {author} {\bibfnamefont {M.}~\bibnamefont
  {Lostaglio}},\ }\href {\doibase 10.1103/PRXQuantum.1.010309} {\bibfield
  {journal} {\bibinfo  {journal} {PRX Quantum}\ }\textbf {\bibinfo {volume}
  {1}},\ \bibinfo {pages} {010309} (\bibinfo {year} {2020})}\BibitemShut
  {NoStop}%
\bibitem [{\citenamefont {Nazarov}\ and\ \citenamefont
  {Kindermann}(2003)}]{NazarovEURPHYS2003}%
  \BibitemOpen
  \bibfield  {author} {\bibinfo {author} {\bibfnamefont {Y.}~\bibnamefont
  {Nazarov}}\ and\ \bibinfo {author} {\bibfnamefont {M.}~\bibnamefont
  {Kindermann}},\ }\href {https://doi.org/10.1140/epjb/e2003-00293-1}
  {\bibfield  {journal} {\bibinfo  {journal} {Eur. Phys. J. B}\ }\textbf
  {\bibinfo {volume} {35}},\ \bibinfo {pages} {413} (\bibinfo {year}
  {2003})}\BibitemShut {NoStop}%
\bibitem [{\citenamefont {Clerk}(2011)}]{ClerkPRA2011}%
  \BibitemOpen
  \bibfield  {author} {\bibinfo {author} {\bibfnamefont {A.}~\bibnamefont
  {Clerk}},\ }\href {https://link.aps.org/doi/10.1103/PhysRevA.84.043824}
  {\bibfield  {journal} {\bibinfo  {journal} {Phys. Rev. A}\ }\textbf {\bibinfo
  {volume} {84}},\ \bibinfo {pages} {043824} (\bibinfo {year}
  {2011})}\BibitemShut {NoStop}%
\bibitem [{\citenamefont {Hofer}\ and\ \citenamefont
  {Clerk}(2016)}]{HoferPRL2016}%
  \BibitemOpen
  \bibfield  {author} {\bibinfo {author} {\bibfnamefont {P.}~\bibnamefont
  {Hofer}}\ and\ \bibinfo {author} {\bibfnamefont {A.}~\bibnamefont {Clerk}},\
  }\href {https://link.aps.org/doi/10.1103/PhysRevLett.116.013603} {\bibfield
  {journal} {\bibinfo  {journal} {Phys. Rev. Lett.}\ }\textbf {\bibinfo
  {volume} {116}},\ \bibinfo {pages} {013603} (\bibinfo {year}
  {2016})}\BibitemShut {NoStop}%
\bibitem [{\citenamefont {Gardas}\ and\ \citenamefont
  {Deffner}(2018)}]{gardas2018quantum}%
  \BibitemOpen
  \bibfield  {author} {\bibinfo {author} {\bibfnamefont {B.}~\bibnamefont
  {Gardas}}\ and\ \bibinfo {author} {\bibfnamefont {S.}~\bibnamefont
  {Deffner}},\ }\href@noop {} {\bibfield  {journal} {\bibinfo  {journal}
  {Scientific reports}\ }\textbf {\bibinfo {volume} {8}},\ \bibinfo {pages} {1}
  (\bibinfo {year} {2018})}\BibitemShut {NoStop}%
\bibitem [{\citenamefont {Landi}\ \emph {et~al.}(2020)\citenamefont {Landi},
  \citenamefont {Fonseca~de Oliveira},\ and\ \citenamefont
  {Buksman}}]{PhysRevA.101.042106}%
  \BibitemOpen
  \bibfield  {author} {\bibinfo {author} {\bibfnamefont {G.~T.}\ \bibnamefont
  {Landi}}, \bibinfo {author} {\bibfnamefont {A.~L.}\ \bibnamefont {Fonseca~de
  Oliveira}}, \ and\ \bibinfo {author} {\bibfnamefont {E.}~\bibnamefont
  {Buksman}},\ }\href {\doibase 10.1103/PhysRevA.101.042106} {\bibfield
  {journal} {\bibinfo  {journal} {Phys. Rev. A}\ }\textbf {\bibinfo {volume}
  {101}},\ \bibinfo {pages} {042106} (\bibinfo {year} {2020})}\BibitemShut
  {NoStop}%
\bibitem [{\citenamefont {Buffoni}\ and\ \citenamefont
  {Campisi}(2020)}]{buffoni2020thermodynamics}%
  \BibitemOpen
  \bibfield  {author} {\bibinfo {author} {\bibfnamefont {L.}~\bibnamefont
  {Buffoni}}\ and\ \bibinfo {author} {\bibfnamefont {M.}~\bibnamefont
  {Campisi}},\ }\href@noop {} {\bibfield  {journal} {\bibinfo  {journal}
  {Quantum Science and Technology}\ }\textbf {\bibinfo {volume} {5}},\ \bibinfo
  {pages} {035013} (\bibinfo {year} {2020})}\BibitemShut {NoStop}%
\bibitem [{\citenamefont {Deffner}(2021)}]{deffner2021energetic}%
  \BibitemOpen
  \bibfield  {author} {\bibinfo {author} {\bibfnamefont {S.}~\bibnamefont
  {Deffner}},\ }\href@noop {} {\bibfield  {journal} {\bibinfo  {journal} {arXiv
  preprint arXiv:2102.05118}\ } (\bibinfo {year} {2021})}\BibitemShut {NoStop}%
\bibitem [{\citenamefont {Caruso}\ \emph {et~al.}(2014)\citenamefont {Caruso},
  \citenamefont {Giovannetti}, \citenamefont {Lupo},\ and\ \citenamefont
  {Mancini}}]{Caruso_RevModPhys_2014}%
  \BibitemOpen
  \bibfield  {author} {\bibinfo {author} {\bibfnamefont {F.}~\bibnamefont
  {Caruso}}, \bibinfo {author} {\bibfnamefont {V.}~\bibnamefont {Giovannetti}},
  \bibinfo {author} {\bibfnamefont {C.}~\bibnamefont {Lupo}}, \ and\ \bibinfo
  {author} {\bibfnamefont {S.}~\bibnamefont {Mancini}},\ }\href
  {https://link.aps.org/doi/10.1103/RevModPhys.86.1203} {\bibfield  {journal}
  {\bibinfo  {journal} {Rev. Mod. Phys.}\ }\textbf {\bibinfo {volume} {86}},\
  \bibinfo {pages} {1203} (\bibinfo {year} {2014})}\BibitemShut {NoStop}%
\bibitem [{Note1()}]{Note1}%
  \BibitemOpen
  \bibinfo {note} {Let us observe that, in order to obtain Eq.~\protect \textup
  {\hbox {\mathsurround \z@ \protect \normalfont (\ignorespaces \ref
  {average_DeltaE}\unskip \@@italiccorr )}}, we need to weight the statistics
  of the measurement outcomes at $t=t_{\protect \rm f}$ with the probabilities
  to initially get one of the outcomes $E_{\protect \rm i}$. Otherwise the
  energy variation $\Delta E$ is erroneously proportional to ${\protect \rm
  Tr}[H(t_{\protect \rm f})\rho _{\protect \rm f}]$.}\BibitemShut {Stop}%
\bibitem [{SM()}]{SM}%
  \BibitemOpen
  \href@noop {} {}\bibinfo {note} {Supplementary Material available from {XXXX}
  include additional technical details on the analysis reported in the main
  text.}\BibitemShut {Stop}%
\bibitem [{\citenamefont {Perarnau-Llobet}\ \emph {et~al.}(2017)\citenamefont
  {Perarnau-Llobet}, \citenamefont {B\"{a}umer}, \citenamefont {Hovhannisyan},
  \citenamefont {Huber},\ and\ \citenamefont {Acin}}]{Perarnau-LlobetPRL2017}%
  \BibitemOpen
  \bibfield  {author} {\bibinfo {author} {\bibfnamefont {M.}~\bibnamefont
  {Perarnau-Llobet}}, \bibinfo {author} {\bibfnamefont {E.}~\bibnamefont
  {B\"{a}umer}}, \bibinfo {author} {\bibfnamefont {K.}~\bibnamefont
  {Hovhannisyan}}, \bibinfo {author} {\bibfnamefont {M.}~\bibnamefont {Huber}},
  \ and\ \bibinfo {author} {\bibfnamefont {A.}~\bibnamefont {Acin}},\ }\href
  {https://link.aps.org/doi/10.1103/PhysRevLett.118.070601} {\bibfield
  {journal} {\bibinfo  {journal} {Phys. Rev. Lett.}\ }\textbf {\bibinfo
  {volume} {118}},\ \bibinfo {pages} {070601} (\bibinfo {year}
  {2017})}\BibitemShut {NoStop}%
\bibitem [{Note2()}]{Note2}%
  \BibitemOpen
  \bibinfo {note} {Similarly, the same result holds if we compare the
  probability density function of the EPM protocol, for a general initial state
  this time, with the one of the MLL scheme~\cite {MicadeiPRL2020} (see also
  the SM to this work). We thank Gabriel Landi for pointing out this result in
  relation to the MLL scheme.}\BibitemShut {Stop}%
\bibitem [{\citenamefont {Cimini}\ \emph {et~al.}(2020)\citenamefont {Cimini},
  \citenamefont {Gherardini}, \citenamefont {Barbieri}, \citenamefont
  {Gianani}, \citenamefont {Sbroscia}, \citenamefont {Buffoni}, \citenamefont
  {Paternostro},\ and\ \citenamefont {Caruso}}]{cimini2020experimental}%
  \BibitemOpen
  \bibfield  {author} {\bibinfo {author} {\bibfnamefont {V.}~\bibnamefont
  {Cimini}}, \bibinfo {author} {\bibfnamefont {S.}~\bibnamefont {Gherardini}},
  \bibinfo {author} {\bibfnamefont {M.}~\bibnamefont {Barbieri}}, \bibinfo
  {author} {\bibfnamefont {I.}~\bibnamefont {Gianani}}, \bibinfo {author}
  {\bibfnamefont {M.}~\bibnamefont {Sbroscia}}, \bibinfo {author}
  {\bibfnamefont {L.}~\bibnamefont {Buffoni}}, \bibinfo {author} {\bibfnamefont
  {M.}~\bibnamefont {Paternostro}}, \ and\ \bibinfo {author} {\bibfnamefont
  {F.}~\bibnamefont {Caruso}},\ }\href {\doibase 10.1038/s41534-020-00325-7}
  {\bibfield  {journal} {\bibinfo  {journal} {npj Quantum Information}\
  }\textbf {\bibinfo {volume} {6}},\ \bibinfo {pages} {1} (\bibinfo {year}
  {2020})}\BibitemShut {NoStop}%
\bibitem [{\citenamefont {Deffner}\ \emph {et~al.}(2016)\citenamefont
  {Deffner}, \citenamefont {Paz},\ and\ \citenamefont
  {Zurek}}]{DeffnerPRE2016}%
  \BibitemOpen
  \bibfield  {author} {\bibinfo {author} {\bibfnamefont {S.}~\bibnamefont
  {Deffner}}, \bibinfo {author} {\bibfnamefont {J.}~\bibnamefont {Paz}}, \ and\
  \bibinfo {author} {\bibfnamefont {W.}~\bibnamefont {Zurek}},\ }\href
  {https://link.aps.org/doi/10.1103/PhysRevE.94.010103} {\bibfield  {journal}
  {\bibinfo  {journal} {Phys. Rev. E}\ }\textbf {\bibinfo {volume} {94}},\
  \bibinfo {pages} {010103(R)} (\bibinfo {year} {2016})}\BibitemShut {NoStop}%
\bibitem [{\citenamefont {Sone}\ \emph {et~al.}(2020)\citenamefont {Sone},
  \citenamefont {Liu},\ and\ \citenamefont {Cappellaro}}]{Sone2020}%
  \BibitemOpen
  \bibfield  {author} {\bibinfo {author} {\bibfnamefont {A.}~\bibnamefont
  {Sone}}, \bibinfo {author} {\bibfnamefont {Y.-X.}\ \bibnamefont {Liu}}, \
  and\ \bibinfo {author} {\bibfnamefont {P.}~\bibnamefont {Cappellaro}},\
  }\href {\doibase 10.1103/PhysRevLett.125.060602} {\bibfield  {journal}
  {\bibinfo  {journal} {Phys. Rev. Lett.}\ }\textbf {\bibinfo {volume} {125}},\
  \bibinfo {pages} {060602} (\bibinfo {year} {2020})}\BibitemShut {NoStop}%
\end{thebibliography}%

\end{document}


\title{Supplemental Materials: The role of quantum coherence in energy fluctuations}

\author{S. Gherardini}\thanks{These authors contributed equally to this work}
\affiliation{CNR-INO \& LENS, via G.  Sansone 1, I-50019 Sesto Fiorentino, Italy.}
\affiliation{\mbox{Department of Physics and Astronomy, University of Florence,} via G.  Sansone 1, I-50019 Sesto Fiorentino, Italy.}
\affiliation{CNR-IOM DEMOCRITOS Simulation Center and SISSA, Via Bonomea 265, I-34136 Trieste, Italy}
\author{A. Belenchia}\thanks{These authors contributed equally to this work}
\affiliation{Institut f\"{u}r Theoretische Physik, Eberhard-Karls-Universit\"{a}t T\"{u}bingen, 72076 T\"{u}bingen, Germany}
\affiliation{Centre for Theoretical Atomic, Molecular and Optical Physics, School of Mathematics and Physics, Queen's University Belfast, Belfast BT7 1NN, United Kingdom}
\author{M. Paternostro}
\affiliation{Centre for Theoretical Atomic, Molecular and Optical Physics, School of Mathematics and Physics, Queen's University Belfast, Belfast BT7 1NN, United Kingdom}
\author{A. Trombettoni}

\affiliation{Department of Physics, University of Trieste, Strada Costiera 11, I-34151 Trieste, Italy}
\affiliation{CNR-IOM DEMOCRITOS Simulation Center and SISSA, Via Bonomea 265, I-34136 Trieste, Italy}

\maketitle
\setcounter{equation}{0}
\setcounter{figure}{0}
\setcounter{table}{0}
\setcounter{page}{1}
\renewcommand{\theequation}{S\arabic{equation}}
\renewcommand{\thefigure}{S\arabic{figure}}
\renewcommand{\bibnumfmt}[1]{[S#1]}
\renewcommand{\citenumfont}[1]{S#1}

\section{Classical uncertainty on the initial state}

The operational protocol that we are introducing in this paper does not reproduce the same results of the two-point measurement (TPM) scheme even in the absence of coherence in the initial state $\rho_{\rm i}$. There is indeed a discrepancy originating from a classical uncertainty on even diagonal (in the initial energy basis) $\rho_{\rm i}$ that is retained in our scheme. Despite this aspect is in agreement with the theses of the no-go theorem~\cite{Perarnau-LlobetPRL2017_SM} as explained in the main text, it is worth understanding it in more detail. In this regard, let us now substitute the density operator
$\varrho \equiv \sum_{r}p_{\rm i}^{(r)}\rho_{\rm i}^{(r)} = \sum_{r}p_{\rm i}^{(r)}\Pi_{\rm i}^{(r)}$ (mixed quantum state diagonal in the energy basis of the system at $t_{\rm i}$, i.e., $[\varrho,H(t_{\rm i})]=0$) as input quantum state $\rho_{\rm i}$ in Eq.~(2) of the main text. One finds that
\begin{eqnarray}\label{eq:eq_1}
{\rm P}_{\rm coh}(\Delta E) &=& \sum_{k,\ell}p_{\rm i}^{(\ell)}p_{\rm f}^{(k)}\delta(\Delta E - \Delta E_{k,\ell})
= \sum_{k,\ell}{\rm Tr}(\Pi_{\rm i}^{(\ell)}\rho_{\rm i}){\rm Tr}(\Pi_{\rm f}^{(k)}\Phi_{t_{\rm f}}[\rho_{\rm i}])\delta(\Delta E - \Delta E_{k,\ell})\nonumber \\
&=& \sum_{k,\ell,r_{1},r_{2}}p_{\rm i}^{(r_1)}p_{\rm i}^{(r_2)}{\rm Tr}(\Pi_{\rm i}^{(\ell)}\Pi_{\rm i}^{(r_1)}){\rm Tr}(\Pi_{\rm f}^{(k)}\Phi_{t_{\rm f}}[\Pi_{\rm i}^{(r_2)}])\delta(\Delta E - \Delta E_{k,\ell})\nonumber \\
&=& \sum_{k,r_{1},r_{2}}p_{\rm i}^{(r_1)}p_{\rm i}^{(r_2)}{\rm Tr}(\Pi_{\rm f}^{(k)}\Phi_{t_{\rm f}}[\Pi_{\rm i}^{(r_2)}])\delta(\Delta E - \Delta E_{k,r_{1}})
= \sum_{k,r_{1},r_{2}}p_{\rm i}^{(r_{1})}p_{\rm f,i}^{(k,r_{2})}\delta(\Delta E - \Delta E_{k,r_{1}}),
\end{eqnarray}
where we have used the relations
${\rm Tr}(\Pi_{\rm i}^{(\ell)}\Pi_{\rm i}^{(r_1)}) = \delta(\ell-r_{1})$ and
$p_{\rm f,i}^{(k,r_{2})} \equiv p_{\rm i}^{(r_2)}{\rm Tr}(\Pi_{\rm f}^{(k)}\Phi_{t_{\rm f}}[\Pi_{\rm i}^{(r_2)}]) = p_{\rm i}^{(r_2)}p_{\rm f|i}^{(k,r_2)}$
with $p_{\rm f,i}^{(k,r_{2})}$ joint probabilities.

From Eq.~(\ref{eq:eq_1}) one can deduce that
${\rm P}_{\rm coh}(\Delta E) = {\rm P}_{\rm TPM}(\Delta E)$ if and only if the initial state is chosen as one of the eigenstates of the initial Hamiltonian, such that $\Delta E_{k,r_{1}} = \Delta E_{k,r_{2}}$. Indeed, in such a case
\begin{equation}\label{eq:eq_2}
{\rm P}_{\rm coh}(\Delta E) = \sum_{r_1}p_{\rm i}^{(r_1)}\sum_{k,r_{2}}p_{\rm f,i}^{(k,r_{2})}\delta(\Delta E - \Delta E_{k,r_{2}})
= \sum_{r_1}p_{\rm i}^{(r_1)}{\rm P}_{\rm TPM}(\Delta E) = {\rm P}_{\rm TPM}(\Delta E) \ .
\end{equation}

It is then clear that an initial uncertainty on which eigenstate of the Hamiltonian needs to be propagated, due to the fact that in the EPM protocol the initial measurement is \textit{virtual}, determines an additional uncertainty on the energy statistics,
which is reflected in the discrepancy between the two methods. The latter is provided by the arbitrariness of the inequality $\Delta E_{k,r_1}\neq\Delta E_{k,r_2}$, which is due to the lack of initial projective measurement and makes it impossible to reduce the two summations in Eq.~\eqref{eq:eq_1} to a single one as in the TPM case.

\section{Recovering the TPM statistics}
\label{sec:app_rec_class_limit}

As stated before, the energy change probability distribution ${\rm P}_{\rm coh}$ does not reduce to the one from the TPM scheme unless the initial state of both protocol is an energy eigenstate. Considering again an initial state diagonal in the energy eigenbasis, it can be easily seen that, to find the same statistics of $\Delta E$ as given by a TPM protocol, Eq.~(2) of the main text has to be employed as many times as the number of probabilities $p_{\rm i}^{(r)}$ that define the initial density operator $\varrho=\sum_{r}p_{\rm i}^{(r)}\Pi_{\rm i}^{(r)}$, each time initializing the quantum system in one of the projectors $\Pi_{\rm i}^{(r)}$. In doing this, the corresponding probability distribution of $\Delta E$ turns out to be
\begin{eqnarray}\label{conventional_prob_E}
  &{\rm P}_{\rm coh}(\Delta E) = \displaystyle{\sum_{r}p_{\rm i}^{(r)}\sum_{k,\ell}\delta(\Delta E - \Delta E_{k,\ell})\delta(\ell-r){\rm Tr}(\Phi_{t_{\rm f}}[\Pi_{\rm i}^{(r)}]\Pi_{\rm f}^{(k)})}&\nonumber \\
  &= \displaystyle{\sum_{k,r}\delta(\Delta E - \Delta E_{k,r})p_{\rm f|i}^{(k,r)}p_{\rm i}^{(r)}} \equiv {\rm P}_{\rm TPM}(\Delta E) \,&
\end{eqnarray}
where $p_{\rm f|i}^{(k,r)} \equiv {\rm Tr}(\Phi_{t_{\rm f}}[\Pi_{\rm i}^{(r)}]\Pi_{\rm f}^{(k)})$ is the transition probability to measure the final energy $E^{(k)}_{\rm f}$ conditioned to have obtained $E^{(r)}_{\rm i}$ at $t=t_{\rm i}$. Only in this way, the proposed formalism falls
into the category of FT protocols~\cite{LostaglioPRL2018_SM}, so that we can recover the conventional statistics of energy change as provided by the TPM scheme. This result is not surprising, since we are now analyzing a situation in which a possible first energy measurement at $t=t_{\rm i}$ would not introduce any disturbance to the evolution of the system. As a further remark, also notice that with this approach the notion of quasi-probabilities is not directly used~\cite{AllahverdyanPRE2014_SM,LevyArxiv2019_SM}.

\section{Analysis of the $1$st and $2$nd energy statistical moments}

In this section, we provide the analytical expressions of the $1$st and $2$nd statistical moments of the proposed energy change distribution in comparison with the ones obtained by the TPM protocol and the Micadei-Landi-Lutz (MLL) protocol in Ref.~\cite{Micadei_SM}. In doing this, we recall that the initial state $\rho_{\rm i}$ in Ref.~\cite{Micadei_SM} is expressed in terms of its eigenstates with notation $\sum_{s}p^{(s)}|s\rangle\!\langle s|$, which is the same that we will use in the following. We list below all the formulas of the joint probability $p(E_{\rm i}^{(\ell)},E_{\rm f}^{(k)})$
and the $1$st and $2$nd statistical moments of $\Delta E$ that one can obtain from the three methods.\\ \\
\textbf{EPM protocol proposed in the present paper:}
\begin{eqnarray}
    && p(E_{\rm i}^{(\ell)},E_{\rm f}^{(k)}) = {\rm Tr}(\rho_{\rm i}\Pi_{\rm i}^{(\ell)})\,{\rm Tr}(\Phi_{t_{\rm f}}[\rho_{\rm i}]\Pi_{\rm f}^{(k)}) \\
    && \langle\Delta E\rangle = {\rm Tr}({H}(t_{\rm f})\Phi_{t_{\rm f}}[\rho_{\rm i}]) - {\rm Tr}(H(t_{\rm i})\rho_{\rm i}) \\
    && \langle\Delta E^2\rangle = {\rm Tr}({H}^2(t_{\rm i})\rho_{\rm i}) + {\rm Tr}({H}^2(t_{\rm f})\Phi_{t_{\rm f}}[\rho_{\rm i}]) - 2\,{\rm Tr}(\Phi_{t_{\rm f}}[\rho_{\rm i}]{H}(t_{\rm f}))\,{\rm Tr}(\rho_{\rm i}{H}(t_{\rm i})) \ . \label{eq:app_2nd_moment}
\end{eqnarray}\\
\textbf{MLL protocol:}
\begin{eqnarray}
    && p(E_{\rm i}^{(\ell)},E_{\rm f}^{(k)}) = \sum_s p^{(s)}{\rm Tr}(|s\rangle\!\langle s|\Pi_{\rm i}^{(\ell)})\,{\rm Tr}(\Phi_{t_{\rm f}}[|s\rangle\!\langle s|]\Pi_{\rm f}^{(k)}) \\
    && \langle\Delta E\rangle = {\rm Tr}({H}(t_{\rm f})\Phi_{t_{\rm f}}[\rho_{\rm i}]) - {\rm Tr}({H}(t_{\rm i})\rho_{\rm i}) \\
    &&  \langle\Delta E^2\rangle = {\rm Tr}({H}^2(t_{\rm i})\rho_{\rm i}) + {\rm Tr}({H}^2(t_{\rm f})\Phi_{t_{\rm f}}[\rho_{\rm i}]) - 2\,\sum_s p^{(s)}{\rm Tr}(\Phi_{t_{\rm f}}[|s\rangle\!\langle s|]{H}(t_{\rm f}))\,{\rm Tr}(|s\rangle\!\langle s|{H}(t_{\rm i})) \ . \label{eq:app_2nd_moment_MLL}
\end{eqnarray}\\
\textbf{TPM protocol:}
\begin{eqnarray}
    && p(E_{\rm i}^{(\ell)},E_{\rm f}^{(k)}) = {\rm Tr}(\rho_{\rm i}\Pi_{\rm i}^{(\ell)})\,{\rm Tr}(\Phi_{t_{\rm f}}[\Pi_{\rm i}^{(\ell)}]\Pi_{\rm f}^{(k)}) \\
    && \langle\Delta E\rangle = {\rm Tr}\left(H(t_{\rm f})\Phi_{t_{\rm f}}\left[\sum_{\ell}{\rm Tr}(\rho_{\rm i}\Pi_{\rm i}^{(\ell)})\Pi_{\rm i}^{(\ell)}\right]\right) - {\rm Tr}(H(t_{\rm i})\rho_{\rm i}) \\
    &&  \langle\Delta E^2\rangle = {\rm Tr}(H^2(t_{\rm i})\rho_{\rm i}) + 
    {\rm Tr}\left({H}^2(t_{\rm f})\Phi_{t_{\rm f}}\left[\sum_{\ell}{\rm Tr}(\rho_{\rm i}\Pi_{\rm i}^{(\ell)})\Pi_{\rm i}^{(\ell)}\right]\right) - 2\,\sum_{\ell}E_{\rm i}^{(\ell)}{\rm Tr}({H}(t_{\rm f})\Phi_{t_{\rm f}}[\Pi_{\rm i}^{(\ell)}])\,{\rm Tr}(\rho_{\rm i}\Pi_{\rm i}^{(\ell)}) \ .
\end{eqnarray}

Within the TPM protocol, an initial measurement of the system Hamiltonian at $t=t_{\rm i}$
and a final one at $t=t_{\rm f}$ on the conditional evolved states are performed.
In order to get the corresponding conditional probability, the system has to be
separately initialized in each eigenstate of ${H}(t_{\rm i})$, respectively.

Concerning the MLL protocol, for the sake of experimentally characterise the energy change
probability distribution, one needs to initialize the system in the eigenstates
of the initial density matrix $\rho_{\rm i}$. This operation could be equivalently carried
on by performing an initial measurement of the observable
$\mathcal{O}\equiv\sum_{s}o_{s}|s\rangle\!\langle s|$, in general not
commuting with ${H}(t_{\rm i})$. Indeed, according to the MLL protocol,
the final energy measurement is performed on the evolved eigenstate $|s\rangle\!\langle s|$ of the initial state and the results are then
weighted with the probabilities $\{p^{(s)}\}$.

Finally, in our EPM protocol, the initial state $\rho_{\rm i}$ is arbitrary
and the final energy measurement is performed on the evolved initial state
without any need to initialize the system in a different state.
The energy change probability distribution is obtained by weighting these
final probabilities with the ones concerning the initial virtual energy measurement,
which are accessible from the knowledge of the initial state. 

One can observe that the average energy change $\langle\Delta E\rangle$
provided by the EPM protocol and the MLL protocol are the same, differently to
the one from the TPM protocol for which the mean final energy measured at $t=t_{\rm f}$
does not contain any contributions from initial coherence terms in $\rho_{\rm i}$.
Furthermore, regarding the $2$nd moment $\langle\Delta E^{2}\rangle$, the three protocols
differ again for the way in which the initial energy outcomes
(eigenvalues of ${H}(t_{\rm i})$) are taken into account in
relation to $\rho_{\rm i}$. Only our (operational) method makes no assumptions
about $\rho_{\rm i}$, since we completely remove the need to perform
any initial projective measurement. 
However, in general, if the initial state $\rho_{\rm i}$ is pure,
the second moments~\eqref{eq:app_2nd_moment} and \eqref{eq:app_2nd_moment_MLL} coincide,
while, as shown above in 
Section II, $\eqref{eq:app_2nd_moment}$ coincides with the second moment
obtained by applying the TPM protocol for an initial state corresponding to an
eigenstate of ${H}(t_{\rm i})$.

It is also interesting to note that the probability distribution from the EPM protocol corresponds to the product of the marginals of the MLL-protocol probability distribution~\footnote{The authors thank Gabriel T. Landi for pointing this out to us.}.
In particular, the (informational) price that we have to pay due to \textit{not} performing any initial measurement, with respect to the MLL protocol that requires a greater knowledge of the state and dynamics, can be quantified by the mutual
information between the two probability distributions, i.e.,
\begin{equation}
    \mathcal{I}({\rm P}_{\rm MLL},{\rm P}_{\rm coh})=\sum_{k,\ell}p_{\rm MLL}^{(k,\ell)}\log\left(p_{\rm MLL}^{(k,\ell)}/p_{\rm coh}^{(k,\ell)}\right).
\end{equation}
$\mathcal{I}({\rm P}_{\rm MLL},{\rm P}_{\rm coh})$ encodes the cost of our assumption of the statistical independence between the final energy measurement
and the initial virtual one with respect to the MLL scheme.
\\ \\
Let us summarize what we have discussed so far concerning the connection of the proposed protocol with the TPM and MLL schemes:
\begin{itemize}
    \item 
    For an initial state $\rho_{\rm i}$ diagonal in the energy eigenbasis, the MLL and TPM protocols provide the same joint probability $p(E_{\rm i}^{(\ell)},E_{\rm f}^{(k)})$, while the EPM protocol's probability distribution differ from them.
    \item 
    For an initial pure state, not necessarily an eigenstate of the initial Hamiltonian, the joint probabilities from our method and the MLL protocol coincide.
    \item 
    In the special case of initial pure energy eigenstate, all three protocols give the same result.
\end{itemize}
A first element of difference between the EPM protocol and the TPM and MLL ones
is given by a classical uncertainty on the initial state $\rho_{\rm i}$.
This is due to the fact that we are assuming to not know the single pure
components that decompose the initial state $\rho_{\rm i}$, or at least the effect
of the dynamics on them separately. Operationally, both the TPM (explicitly) and the
MLL (implicitly) need to assume the knowledge about the evolution of the pure components
of the system initial state (either in the energy eigenbasis or in its eigenbasis),
which are then evolved and give rise to conditional probabilities. The proposed protocol
does not assume this knowledge and it is thus nicely amenable
for experimental implementations with minimal resources.

\section{Energy change characteristic function}
Here, we provide the mathematical details for the derivation of the characteristic
function associated to the energy change distribution both from the proposed method and the
MLL and TPM protocols. 
The characteristic function from the three methods are respectively equal to 
\begin{align}
    & \mathcal{G}(u) = {\rm Tr}(e^{-iu{H}(t_{\rm i})}\rho_{\rm i})\,{\rm Tr}(e^{iu{H}(t_{\rm f})}\Phi_{t_{\rm f}}[\rho_{\rm i}])\label{eq:app_G_coh} \\
    & \mathcal{G}_{\rm MLL}(u) = \sum_s p^{(s)} {\rm Tr}\left(|s\rangle\!\langle s| e^{-iu{H}(t_{\rm i})}\right)\,{\rm Tr}\left(\Phi_{t_{\rm f}}[|s\rangle\!\langle s|]e^{iu{H}(t_{\rm f})}\right) \\
    & \mathcal{G}_{\rm TPM}(u)={\rm Tr}(e^{iuH(t_{\rm f})}\Phi_{t_{\rm f}}[e^{-iuH(t_{\rm i})}\mathcal{P}])
\end{align}
with $u\in\mathbb{C}$ complex number {and $\mathcal{P}$ the diagonal part of the initial state in the energy basis}. Also at the level of the characteristic function of
the energy change distribution, we can single out coherence contributions. In particular,
by taking $\rho_i=\mathcal{P}+\chi$ in~\eqref{eq:app_G_coh}, one has
\begin{align}
    \mathcal{G}(u) &= {\rm Tr}(e^{-iu{H}(t_{\rm i})}\rho_{\rm i})\,{\rm Tr}(e^{iu{H}(t_{\rm f})}\Phi_{t_{\rm f}}[\rho_{\rm i}])\nonumber \\
    &={\rm Tr}(e^{-iu{H}(t_{\rm i})}\mathcal{P})\,{\rm Tr}(e^{iu{H}(t_{\rm f})}\Phi_{t_{\rm f}}[\mathcal{P}])+{\rm Tr}(e^{-iu{H}(t_{\rm i})}\mathcal{P})\,{\rm Tr}(e^{iu{H}(t_{\rm f})}\Phi_{t_{\rm f}}[\chi])\nonumber \\
    &\equiv\mathcal{G}_{\mathcal{P}}(u)+\mathcal{G}_{\mathcal{\chi}}(u), 
\end{align}
where $\mathcal{G}_{\mathcal{P}}(u) \equiv {\rm Tr}(e^{-iu{H}(t_{\rm i})}\mathcal{P})\,{\rm Tr}(e^{iu{H}(t_{\rm f})}\Phi_{t_{\rm f}}[\mathcal{P}])$ and
$\mathcal{G}_{\mathcal{\chi}}(u) \equiv {\rm Tr}(e^{-iu{H}(t_{\rm i})}\mathcal{P})\,{\rm Tr}(e^{iu{H}(t_{\rm f})}\Phi_{t_{\rm f}}[\chi])$. As a result,
$\mathcal{G}_{\chi}=0$ when $\chi=0$ and $\mathcal{G}_{\mathcal{P}}=\mathcal{G}_{\rm TPM}$
as far as $\mathcal{P}$ is a projector associated to a system energy eigenspace. 

Given the expression of $\mathcal{G}$ for the EPM protocol,
we present in the following a derivation of Eq.\,(8) of the main text. We have 
\begin{align}
    \langle e^{-\beta\Delta E}\rangle&={\rm Tr}(\rho_{\rm i}e^{\beta H(t_{\rm i})}){\rm Tr}(\Phi_{t_{\rm f}}[\rho_{\rm i}]e^{-\beta H(t_{\rm f})})\nonumber\\
    &={\rm Tr}[\mathcal{P}e^{\beta H(t_{\rm i})}]\left({\rm Tr}\left(\Phi_{t_{\rm f}}[\mathcal{P}]e^{-\beta H(t_{\rm f})}\right)+{\rm Tr}\left(\Phi_{t_{\rm f}}[\chi]e^{-\beta H(t_{\rm f})}\right)\right),
\end{align}
where in the second line we have used $\rho_{\rm i}=\mathcal{P}+\chi$ with ${\rm Tr}(\chi)=0$ and $\mathcal{P}$ the diagonal part of the initial state in the energy basis. Further assuming $\mathcal{P}=\rho_{\rm i}^{\rm th}$ and multiplying by $Z_{\rm i}/Z_{\rm f}$ both sides of the equation above, we obtain
\begin{align}\label{eqJarSI}
    \langle e^{-\beta(\Delta E-\Delta F)}\rangle&={\rm Tr}(\mathbb{I})\left({\rm Tr}\left(\Phi_{t_{\rm f}}[\rho_{\rm i}^{\rm th}]\rho_{\rm f}^{\rm th}\right)+{\rm Tr}\left(\Phi_{t_{\rm f}}[\chi]\rho_{\rm f}^{\rm th})\right)\right)=d\,\left({\rm Tr}\left(\Phi_{t_{\rm f}}[\rho_{\rm i}^{\rm th}]\rho_{\rm f}^{\rm th}\right)+{\rm Tr}\left(\Phi_{t_{\rm f}}[\chi]\rho_{\rm f}^{\rm th})\right)\right)
\end{align}
where $\rho_{\rm i(f)}^{\rm th} \equiv \displaystyle{e^{-\beta {H}(t_{\rm i(f)})}/{Z_{\rm i(f)}}}$ are the two thermal (reference) states at inverse temperature $\beta$ referring, respectively, to the initial and final time instants of the protocol, and we have used the fact that $\Delta F = -\beta^{-1}\ln\left(Z_{\rm f}/{Z_{\rm i}}\right)$. 
We refer to Fig.~\ref{newfig2} below for the comparison between the results provided by Eq.~\eqref{eqJarSI} and those coming from the TPM protocol in the case of an open three-level system with time-dependent Hamiltonian and coupled to three thermal baths, one for each level. Instead, Fig.~\ref{newfig} \textbf{(b)} shows the same comparison for the case in which the three-level system is not coupled to the thermal environment, thus evolving unitarily. 
The main difference between the open and close systems lays in the fact that, for the open quantum case of Fig.~\ref{newfig2}, also the TPM protocol's result deviates from unity. 


\section{Experimental Results}

In order to illustrate the effects of initial coherence on energy fluctuations, as singled out by our EPM protocol, in this section we provide more details on the experiments performed with IBM Quantum Experience (IBMQ) and discussed in the main text. 

\subsection{Two-qubit state initialization}

As discussed in the main text, the initial (pure) state of the two-qubits is prepared to be of the form $\rho_{\rm i} =|\psi_\text{i}\rangle\!\langle\psi_\text{i}| = \rho^{\rm th}+\chi$ where the corresponding diagonal part (in the computational basis of the quantum gate) is a thermal state of the local Hamiltonian $H_A+H_B=\epsilon(\sigma_z^{(A)}\otimes\mathbb{I}^{(B)}+\mathbb{I}^{(A)}\otimes\sigma_z^{(B)})$. The initial state, to be implemented on IBMQ, is pure and can be generated by directly acting on the default state $|00\rangle$ of the information carriers via local gates. We request a pure state whose diagonal part $\rho^\text{th}$ takes the thermal-like form
\begin{equation}
    \rho^{\rm th} =\frac{1}{(\cosh (2 \beta  \epsilon )+1)^2} \rm{diag}\left\{{e^{-2 \beta  \epsilon } \cosh ^2(\beta  \epsilon )},{\cosh ^2(\beta  \epsilon )},{\cosh ^2(\beta  \epsilon )},{e^{2 \beta  \epsilon } \cosh ^2(\beta  \epsilon )}\right\}
\end{equation}
with $\beta$ inverse temperature. As any two-qubit state can be expressed, in the computational basis, as $|\psi\rangle\!\langle\psi|=\sum_{k,\ell=0}^{3}a_{k\ell}|k\rangle\!\langle\ell|$, we require 
\begin{equation}\label{phases}
a_{k\ell}=e^{i\phi_{k\ell}} \sqrt{\rho^{\rm th}_{k\ell,k\ell}}\quad\forall{k,l=0,\ldots,3}.
\end{equation}
The corresponding state $|\psi_{\rm i}\rangle$ can be generated by applying on $|00\rangle$ the unitary gate 
\begin{equation}
V \equiv a_{00}(\mathbb{I}\otimes\mathbb{I}) + a_{01}(\mathbb{I}\otimes\sigma_x^{(B)})+a_{10}(\sigma_x^{(A)}\otimes\mathbb{I})+a_{11}(\sigma_x^{(A)}\otimes \sigma_x^{(B)})
\end{equation}
with $a_{ij}\in\mathbb{C}$, in general. By following Ref.\,\cite{ShendePRA2004}, it is found that $V$ can be implemented by means of just local  gates, so that the initial state $\rho_{\rm i}$ is separable. 

In the experiments we have performed, the initial state $|\psi_{\rm i}\rangle$ is chosen so that $\phi_{k\ell}=0\,\forall{k,\ell}$ and it can be obtained by applying two identical single-qubit rotations 
\begin{equation}
\label{rot}
U(\theta_0)\equiv\left(
\begin{array}{cc}
 \cos \left(\frac{\theta_0}{2}\right) &  -\sin \left(\frac{\theta }{2}\right) \\
  \sin \left(\frac{\theta_0 }{2}\right) & \cos \left(\frac{\theta }{2}\right)  \\
\end{array}
\right)
\end{equation}
with the rotation angle $\theta_0$ that must be chosen depending on the value of the inverse temperature $\beta$ and the frequency $\epsilon$.
More explicitly
\begin{equation}
  \text{sech}(\beta) =  \sin(\theta_0),
\end{equation}
in units of $\epsilon^{-1}$. Eq.~\eqref{rot} is routinely implemented in the IBMQ quantum computer. Given the separable nature of the initial state chosen for the experiments, here we are not investigating the role of the initial entanglement in the energetics of the process but we have limited ourselves to investigate the role of the initial coherence in the computational basis. We leave the investigation of the role of entanglement for future works.

\subsection{Two-qubit gates}

After initialization, the experiments proceed with two initial projective measurements for the TPM scheme and no measurement for the EPM protocol. Afterwards, as shown in the top panel of Fig.\,2 in the main manuscript, the circuit implements a controlled gate, followed by  
a final energy measurement in the computational basis of the two-qubits. The controlled unitary reads 
\begin{equation}
|0\rangle\!\langle0|_A\otimes\mathbb{ I}^{(B)}+|1\rangle\!\langle1|_A\otimes U^{(B)}(\theta,\phi,\lambda)
\end{equation}
with $U(\theta,\phi,\lambda)$ the  
standard $U$-gate provided by IBMQ, i.e.,
\begin{equation}
U^{(B)}(\theta,\phi,\lambda)\equiv\left(
\begin{array}{cc}
 \cos \left(\frac{\theta }{2}\right) & -\exp (i \lambda ) \sin \left(\frac{\theta }{2}\right) \\
 \exp (i \phi ) \sin \left(\frac{\theta }{2}\right) & \cos \left(\frac{\theta }{2}\right) \exp (i (\lambda +\phi )) \\
\end{array}
\right)
\end{equation}
that generalizes the rotation used for the preparation of the initial state of the computer. In our experiments, we have used $U(\theta_n,0,0)$, where $\theta_n$ varies from zero to $2\pi$ in steps of $\pi/10$. 

\subsection{Additional experimental results}

In this subsection we report additional results with respect to what was shown in the main text. In particular, we focus our attention on the experimental values of the energy change statistical moments. We also show how the behaviour of the experimental data is well captured by working at the liner response level for what concerns the characteristic functions. 
In other terms, we observe that the expressions for the characteristic functions of the energy change distribution can be expanded as a function of few statistical moments (just two suffice) with a negligible error. This allows us to just consider the experimental data of the first and second statistical moments of the energy-change distribution to properly describe the energetics of the quantum circuits at hand.

\begin{figure*}[t!]
\centering
\includegraphics[scale=0.6]{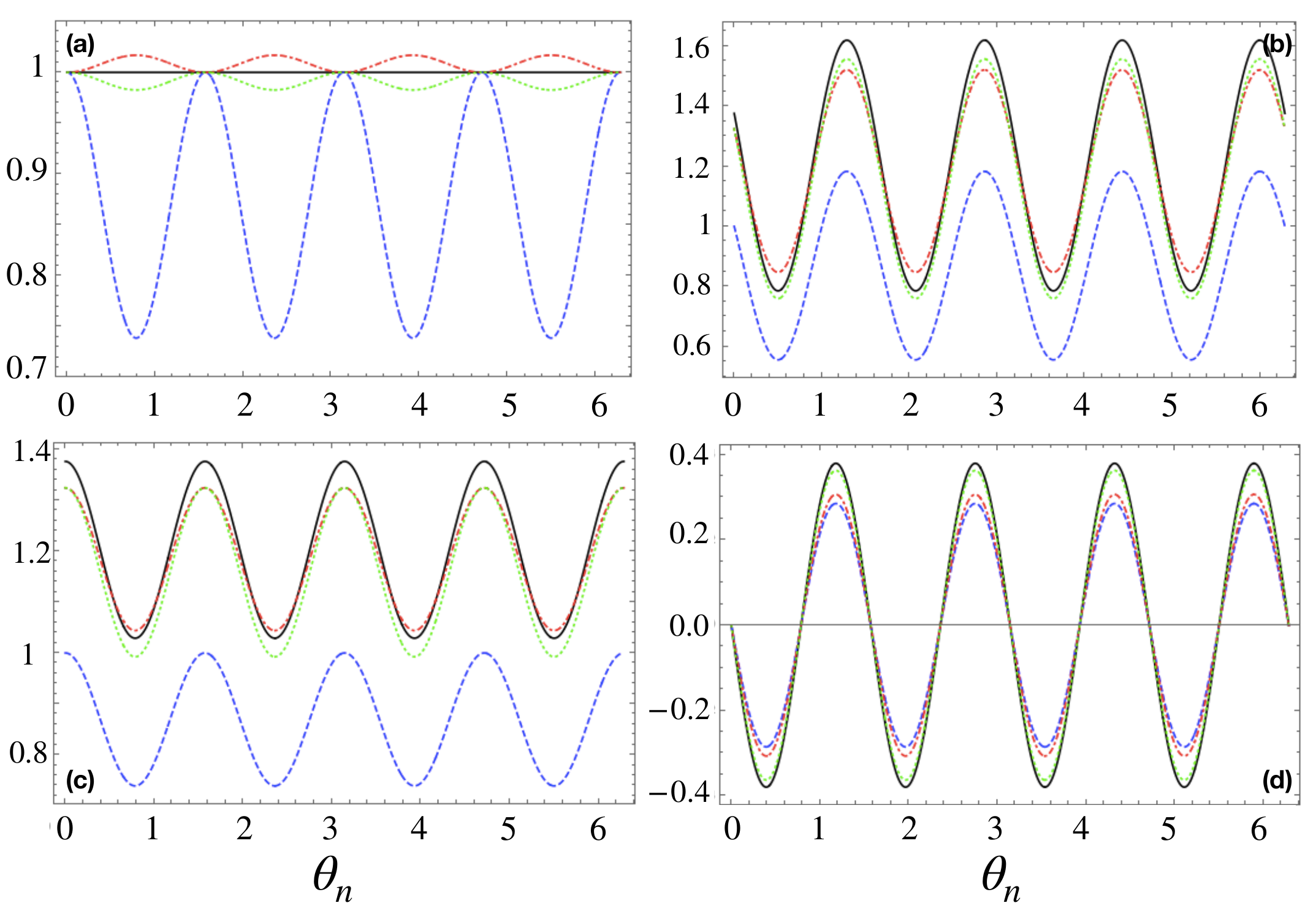}
\caption{Comparison between characteristic functions and the result obtained by only using few statistical moments of the probability distribution. The solid lines represent the characteristic functions $\mathcal{G}(i\beta)$, the blue dashed lines the results obtained retaining only the first moments, the red dot-dashed lines the result obtained keeping until the second moment and the green dotted lines the results obtained keeping until the third moments.\textbf{(a)}: TPM scheme; \textbf{(b)}: EPM scheme; \textbf{(c)}: diagonal contribution to the EPM scheme; \textbf{(d)}: coherence contribution to the EPM scheme. The numerical values of the parameters used and the initial state are the same as for the experiments discussed in the text.}
\label{LR}
\end{figure*}

\begin{figure*}[t!]
\centering
\includegraphics[scale=0.6]{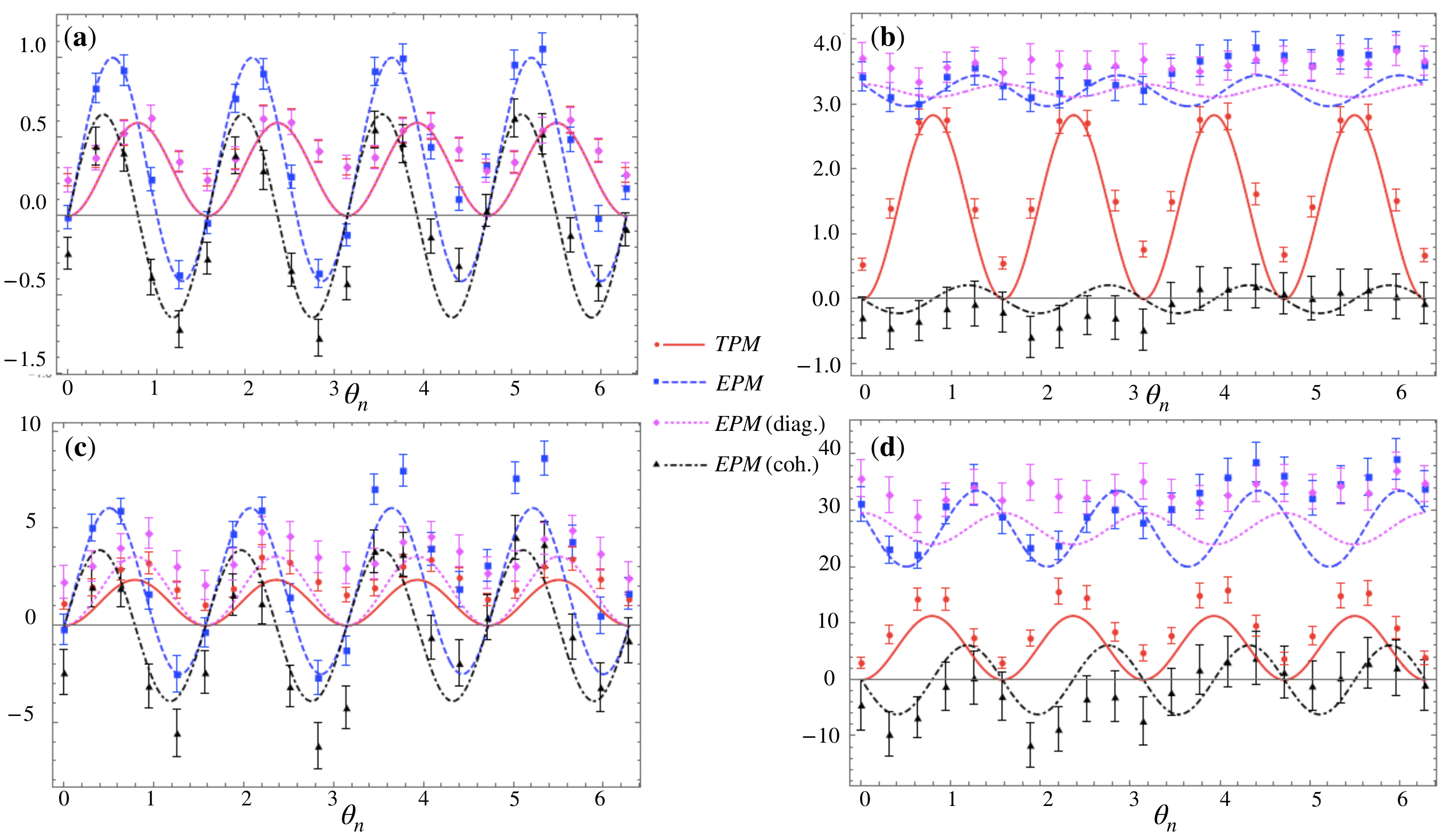}
\caption{Comparison between the theoretical, experimental, and simulated results for the firsts statistical moments of the TPM and EPM statistics. Panels \textbf{(a)} show the first moments of the TPM and EPM scheme, as well as the diagonal and coherence parts of the EPM one. Note that, for the first moments the prediction, and results, of the TPM scheme and the diagonal part of the EPM one are the same. Panels \textbf{(b-d)}: same as before but for the second, third and fourth moments.}
\label{first}
\end{figure*}

For completeness, we report the analytical expressions 
of the energy-chance characteristic functions for an initial pure state of the form $\rho_{\rm i}=\rho^{\rm th}+\chi$, with all the phases in Eq.~\eqref{phases} vanishing. We have
\begin{align}
    &\mathcal{G}_{\rm TPM}(i\beta)=1,\\
    &\mathcal{G}_{\rm EPM}(i\beta)=\frac{4 \left(e^{6 \beta  \epsilon } \left(\sin (2 \theta )-e^{\beta  \epsilon } \cos (2 \theta )\right)^2+e^{4 \beta  \epsilon } \left(e^{\beta  \epsilon } \sin (2 \theta )+\cos (2 \theta )\right)^2+e^{4 \beta  \epsilon }+1\right)}{\left(e^{2 \beta  \epsilon }+1\right)^4},\\
    &\mathcal{G}_{\rm EPM}^{(\rm{diag})}(i\beta)=\frac{4 \left(2 e^{6 \beta  \epsilon } \sin ^2(2 \theta )+\left(e^{4 \beta  \epsilon }+e^{8 \beta  \epsilon }\right) \cos ^2(2 \theta )+e^{4 \beta  \epsilon }+1\right)}{\left(e^{2 \beta  \epsilon }+1\right)^4},\\
    &\mathcal{G}_{\rm EPM}^{(\rm{coh})}(i\beta)=-\frac{1}{2} e^{2 \beta  \epsilon } \sin (4 \theta ) \tanh (\beta  \epsilon ) \text{sech}^3(\beta  \epsilon )\,.
\end{align}
In Fig.~\ref{LR}, we show how considering only up to the third moment of the probability distributions reproduces the main features of  each characteristic function. This leads us to investigate the first moments explicitly, also at the experimental level. The results of this analysis are reported in Fig.~\ref{first} in which we show the first four statistical moments of the energy change probability distributions obtained by applying both the TPM and EMP protocols. In particular, Fig.~\ref{first} shows the comparison of the experimental data for the first to the fourth moments with the analytical curves.

It is worth noticing the growing discrepancy between the analytical results and the experimental ones for higher moments of the probability distributions. In particular, the EPM statistics seems to be particularly affected by this. We can link this effect to the idealization of the system as completely unitary in the theoretical calculations with respect to the actual IBMQ architecture.

\section{Further application: Three-level system in contact with thermal baths}

\begin{figure*}[t!]
\centering
\includegraphics[scale=0.8]{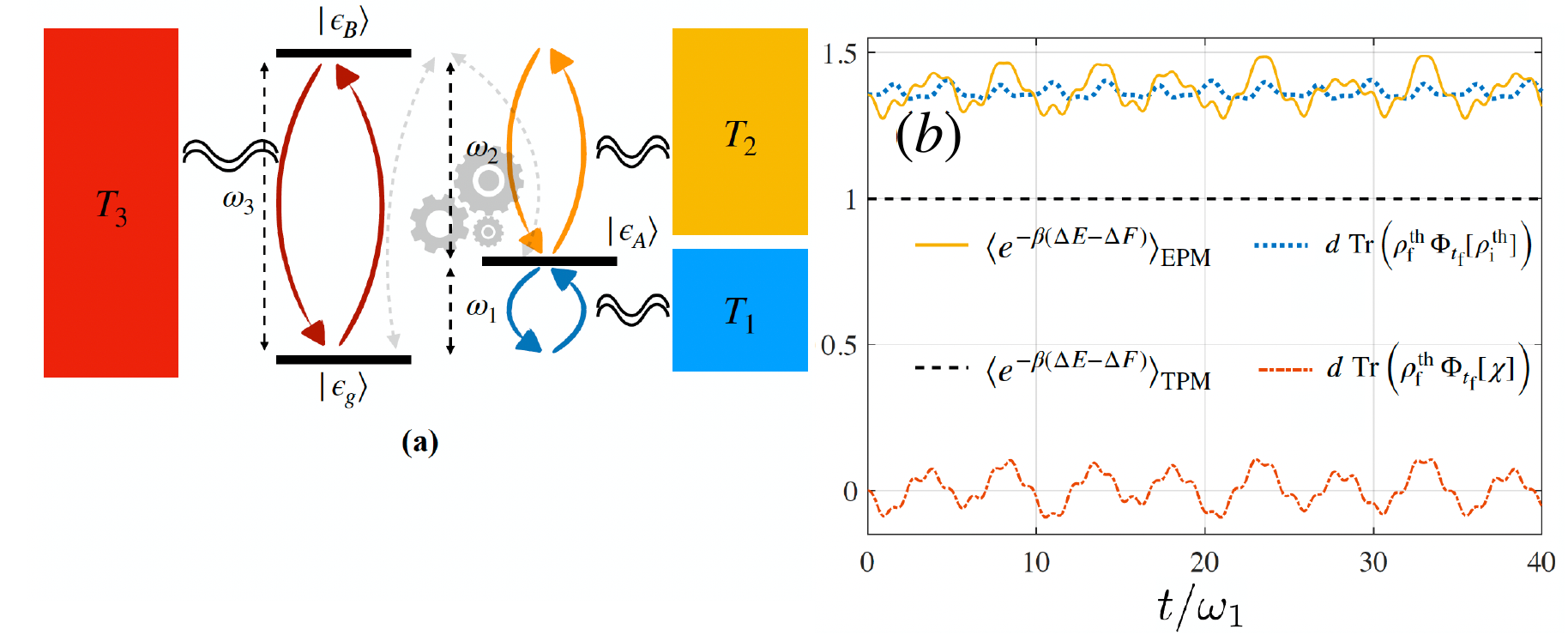}
\caption{
\textbf{(a)}: Pictorial illustration of a $3$-level system coupled to three thermal baths at different temperatures $T_k$, $k=1,2,3$, and externally driven by a time-dependent Hamiltonian term (in light-grey) with {$\omega_2=\omega_3-\omega_1$.} \textbf{(b)}: 
Dashed black line: $\langle e^{-\beta(\Delta E-\Delta F)}\rangle=1$ resulting from the TPM protocol applied to a closed and driven three-level system. Solid yellow curve: $\langle e^{-\beta(\Delta E-\Delta F)}\rangle$ as given by the EPM protocol. Dotted blue curve: $1^{\rm st}$ term on the right hand side of Eq.~\eqref{eqJarSI}. Dash-dotted orange curve: contribution of the initial quantum coherence $\chi$. The initial state is  
$\rho_{\rm i}=\rho^{\rm th}_{\rm i}+\chi$ with $\beta=0.6$ and a random $\chi$ such that $\rho_{\rm i}$ is physical.
}
\label{newfig}
\end{figure*}

In order to showcase once more the effect of initial coherence singled out by our EPM protocol, in this section we address an archetypal quantum thermal machine (see also Refs.~\cite{ScovilPRL59,PalaoPRE2001,KosloffARPC2014}), i.e. a three-level quantum system interacting with thermal reservoirs and driven by a time-dependent Hamiltonian, as shown in Fig.~\ref{newfig} \textbf{(a)}. In contrast with the actual experiments presented in the main text, this example is a numerical exercise which, however, considers a relevant system for quantum thermodynamic considerations and allows us to address in more detail the entropic consequences of adopting the EPM. Furthermore, we also explicitly confront the predictions of the EPM protocol with the ones that can be obtained from the MLL one. 

The dynamics of the system is described by
the master equation 
\begin{equation}\label{me3main}
  \dot{\rho} = -i[H_{\rm t},\rho] + \sum_{i\neq j=1}^{3} \left(L_{ij}\rho L_{ij}^{\dag} - \frac{1}{2}\{L_{ij}^{\dag} L_{ij},\rho\}\right).
\end{equation}
Here, $L_{ij}\equiv\sqrt{\eta_{ij}}|\epsilon_i\rangle\!\langle \epsilon_j|$
is a jump operator acting at rate
$\eta_{ij}$ with
\begin{equation}
\begin{aligned}
    & \eta_{gA}=\gamma\,(n_1^{th}+1), \quad \eta_{Ag}=\gamma\,n_1^{th}, \quad \eta_{AB}=\gamma\,(n_2^{th}+1),\\
    & \eta_{BA}=\gamma\,n_2^{th}, \quad \eta_{gB}=\gamma\,(n_3^{th}+1), \quad \eta_{Bg}=\gamma\,n_3^{th} 
\end{aligned}
\end{equation}
where $n_r^{\rm th}=(e^{\beta_r\omega_r}+1)^{-1}$ and {$\omega_2=\omega_3-\omega_1$}.
The Hamiltonian of the system can be written as $H_{\rm t}=H+H_{\rm{drive}}(t)$ where the free-system Hamiltonian is given by
\begin{equation}
    H=\omega_3|\epsilon_B\rangle\!\langle\epsilon_B|+\omega_1 |\epsilon_A\rangle\!\langle\epsilon_A| 
\end{equation}
with its eigensystem $\{\ket{\epsilon_g},\ket{\epsilon_A},\ket{\epsilon_B};0,\omega_1,\omega_3\}$, while the external driving term is represented by the following time-dependent Hamiltonian
\begin{equation}\label{SMdrive}
    H_{\rm{drive}}(t)=g(t)(|\epsilon_g\rangle\!\langle\epsilon_B|+{\rm h.c.})+f(t)(|\epsilon_A\rangle\!\langle\epsilon_B|+{\rm h.c.})
\end{equation}
that drives the transitions between the second excited state and both the ground and first-excited states with $f(t)$ and $g(t)$ time-dependent rates. 

Note that, for reservoirs at the same temperature and $H_{\rm drive}=0$, the system relaxes to $\rho^\text{th}_{\infty} \equiv e^{-\beta H}/{\rm Tr}(e^{-\beta H})$: independently of $\rho_{\rm i}$, the distribution of the EPM scheme converges to the one resulting from MLL and TPM. This holds for any map with a unique fixed-point. Differently, at finite times the statistics from the three approaches differ even for no coherence in $\rho_{\rm i}$. 

We first address the unitary case, i.e., when the three-level system is decoupled from the thermal reservoirs. In this case, the TPM protocol leads to the Jarzynski equality $\langle e^{-\beta(\Delta E-\Delta F)}\rangle=1$. This is compared to the contribution in Eq.~\eqref{eqJarSI} (Eq.(8) of the main text)
linked to the diagonal and off-diagonal parts of the initial state $\rho_{\rm i}$ in Fig.~\ref{newfig} {\bf (b)}.

\begin{figure*}[t!]
\centering
\includegraphics[scale=0.95]{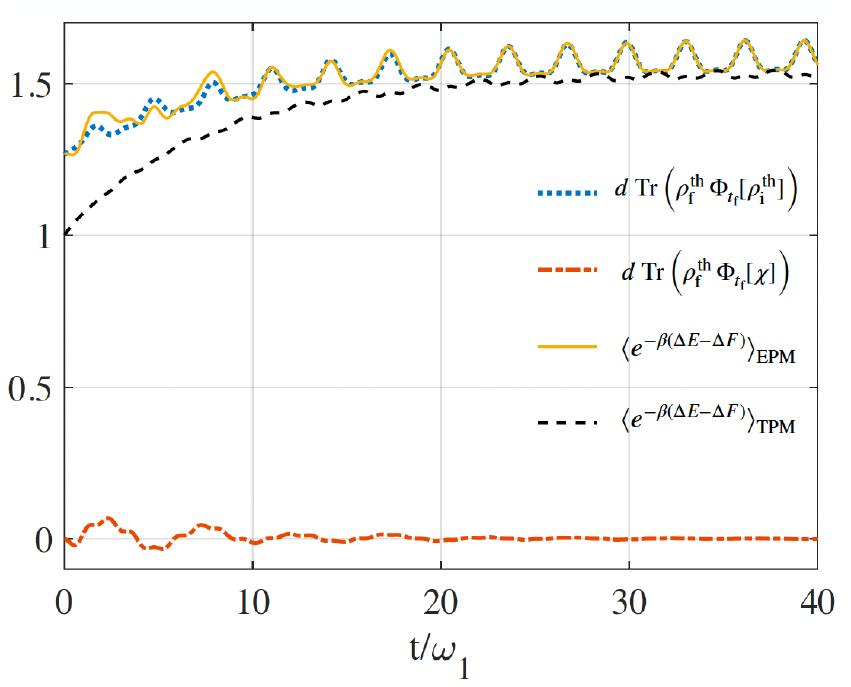}
\caption{Comparison between the quantum Jarzynski-like identities as obtained from the EPM and TPM protocols, respectively [cf. Eqs.~(8) of the main text and \eqref{eqJarSI}]. The dashed black curve shows $\langle e^{-\beta(\Delta E-\Delta F)}\rangle_{\rm TPM}$ resulting from the use of the TPM protocol, the solid yellow curve is for the results achieved using the EPM protocol, while the dotted blue curve is for the first term on the right hand side of Eq.~\eqref{eqJarSI}. The dash-dotted orange curve represents the contribution depending only on the initial quantum coherence brought forward by $\rho_{\rm i}$. Note that, in obtaining these curves, the driving term in Eq.~\eqref{SMdrive} has been included and the initial state as been chosen as $\rho_{\rm i}=\exp{\left(-\beta H(t_{\rm i})\right)/Z_{\rm i}+\chi}$ with $\beta=0.5$ and $\chi$ being randomly generated, but such that $\rho_{\rm i}$ is a legitimate physical state.}
\label{newfig2}
\end{figure*}

A similar comparison can be done in the case of open quantum system dynamics, when the system interact with the thermal reservoirs. In Fig.~\ref{newfig2}, we consider the deviation of $\langle e^{-\beta(\Delta E-\Delta F)}\rangle_{\rm EPM}$ from the result of the TPM scheme, as encoded in Eq.~\eqref{eqJarSI} (namely Eq.~(8) of the main text), and we also plot the two contributions on the right hand side of Eq.~\eqref{eqJarSI} linked to the diagonal and off-diagonal parts of the initial state $\rho_{\rm i}$ respectively. Note that in this case, also the TPM result differ from unity being the open system dynamics neither unitary nor unital.
\begin{figure}[t!]
\centering
\includegraphics[scale=0.9]{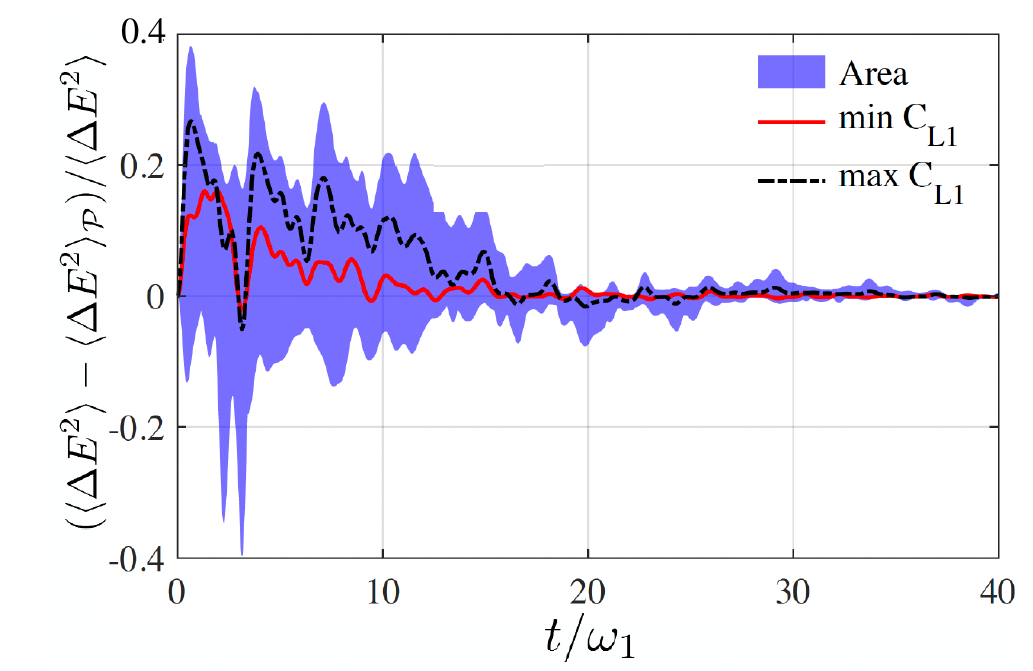}
\caption{$1 - \langle\Delta E^2\rangle_{\mathcal{P}}/\langle\Delta E^2\rangle$ against time for Eq.~\eqref{me3main}
with  $\omega_k = k\omega_1$, $\gamma=0.1\omega_1$, $\beta_1=3$, $\beta_2=1$, $\beta_3=2$, $g(t)=1.5\sin^2(t)$ and $f(t)=1.5-g(t)$ with  $\hbar = k_B = 1$.
}
\label{fig:2nd_moment_coh}
\end{figure}
Fig.~\ref{fig:2nd_moment_coh} shows the temporal behavior of $1 - \langle\Delta E^2\rangle_{\mathcal{P}}/\langle\Delta E^2\rangle$,
evaluated using Eq.~(6) of the main text and
~\eqref{me3main}, thus addressing the contribution
to the $2^\text{nd}$ moment of $\Delta E$ originated by the
initial-coherence terms in $\rho_{\rm i}$. In this case, the contribution of the initial coherence is $\sim40\%$
of the value taken by the $2^\text{nd}$ energy moment. However, while the initial coherences impact significantly on the stochastic energetics, the time behavior of $\langle\Delta E^2\rangle - \langle\Delta E^2\rangle_{\mathcal{P}}$ is not monotonic with the amount of such coherences (as quantified by the measure of quantum coherence $C_{L1} \equiv \frac{1}{2}\sum_{i\neq j}|\rho_{ij}|$~\cite{Baumgratz}).

\begin{figure*}[t!]
\centering
\includegraphics[scale=0.95]{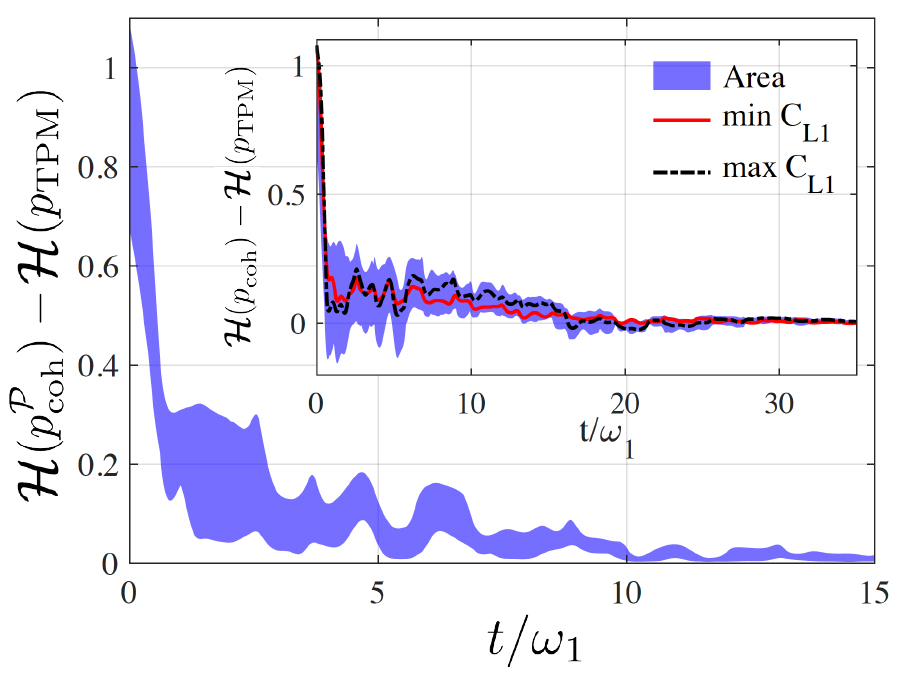}
\caption{
Shannon entropy difference between the EPM and TPM protocols with no coherence in the energy basis. Inset: Difference between the Shannon entropies for the full, non-diagonal, density matrices $\rho_{\rm i}$. In both figures, the blue-shaded regions encompass the values obtained by numerically evaluating the statistics of $\Delta E$ for $10^3$ random initial states --- uniformly sampled by respecting the Haar measure of the space of $3\times 3$ density operators. The red solid lines (black dash-dotted line) denote the corresponding curves obtained by taking in such sample the initial quantum state with the lowest (highest) value of coherence according to the quantum coherence measure $C_{L1}$ as in Ref.~\cite{Baumgratz}. The parameters used in the simulations are $\omega_k = k\omega_1$, $\gamma=0.1\omega_1$, $\beta_1=3$, $\beta_2=1$, $\beta_3=2$, $g(t)=1.5\sin^2(t)$ and $f(t)=1.5(1-\sin^2(2t))$ with  $\omega_1 = \hbar = k_B = 1$.}
\label{fig:2nd_moment_cohSI}
\end{figure*}

In Fig.~\ref{fig:2nd_moment_cohSI} we show the discrepancies between the Shannon entropy of the EPM-based energy change probability distribution in the absence of initial coherence and the one stemming from
TPM-based predictions. As discussed in the main text, this difference quantifies the extra uncertainty, with respect to the TPM scheme, due to not performing the initial energy measurement. Finally, the inset of Fig.~\ref{fig:2nd_moment_cohSI} shows how coherences in the initial state can make the entropy difference negative. This implies that initial coherence could compensate for the extra-uncertainty due to the virtual initial measurement, thus providing a statistically more informative characterisation of energy fluctuations.

From these results, we deduce that the quantum coherence initially present in $\rho_{\rm i}$ (in the system energy basis) has an active role in the first part of the system evolution and is propagated thanks to the action of the driving Hamiltonian. This phenomenon is well-captured by the energy change fluctuations quantified by the EPM protocol. In this specific example, the contribution of the coherence is suppressed at long times. This is due to the fact that the dynamics reaches a (time-dependent) fixed-point, independently of the initial state. Consistently with our previous discussion, in this scenario the EPM probability distribution converges to the TPM one. It should also be noted that, while the initial coherence has a relevant impact on the statistics of the energy fluctuations, also for the Shannon entropy no monotonic relation with the initial coherence is present, as shown in Fig.\,\ref{fig:2nd_moment_cohSI}. The curves corresponding to initial states with maximum coherence never maximize the difference between the results from EPM and TPM distributions, even though they are rather close to it.

Finally, the dynamics of the open quantum system without the external driving and with possibly different temperatures describes processes involving only heat exchanges which are the main focus of Ref.~\cite{Micadei_SM}. Thus, we also explore this case in order to highlight some of the differences between the EPM protocol and the MLL scheme.
\begin{figure}[t]
\centering
\includegraphics[width=0.55\textwidth]{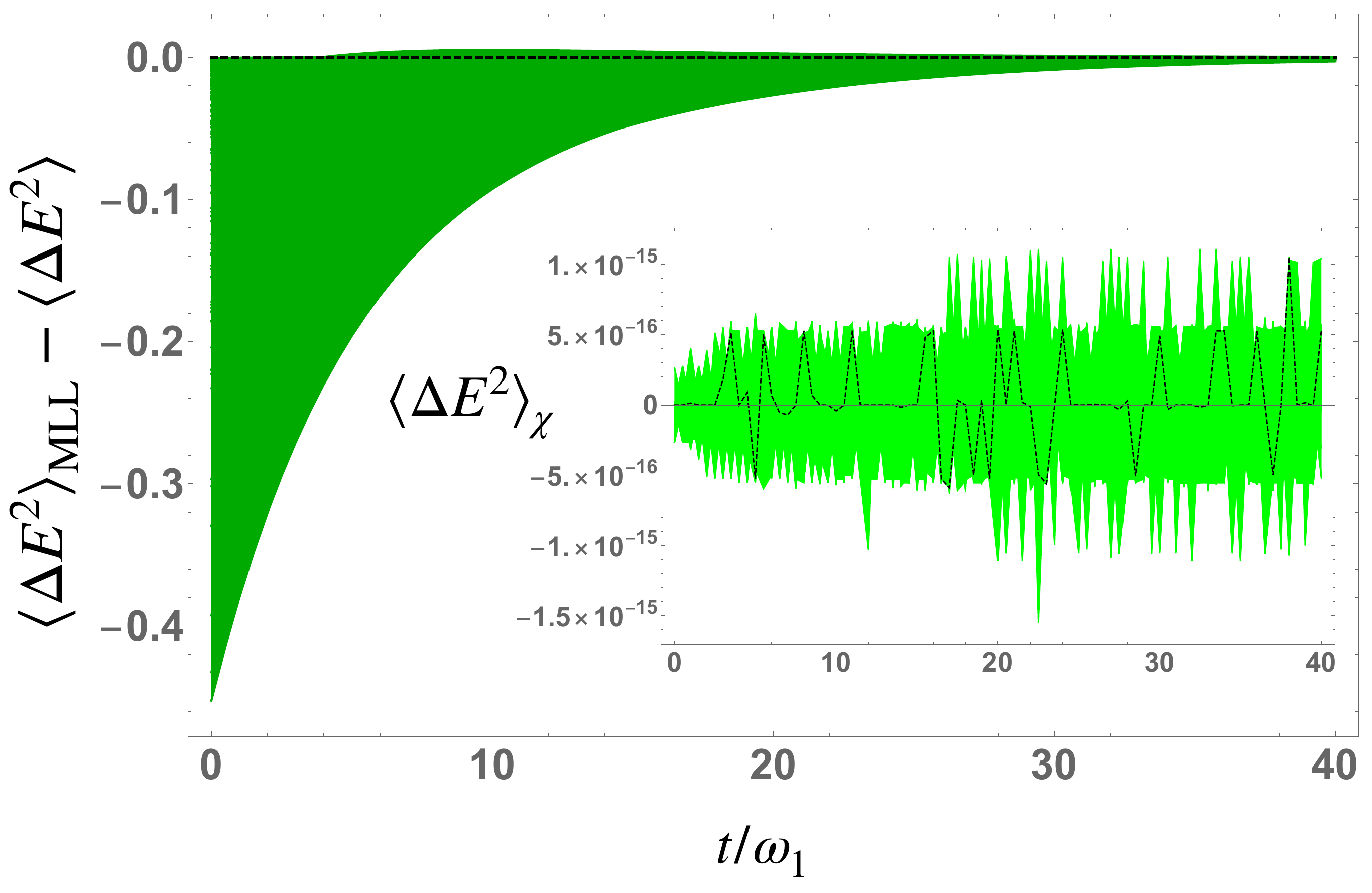}
\caption{Difference between the second moment of
  the MLL probability distribution and the one from the EPM protocol for 100 randomly sampled initial states. The shaded area comprises all the differences, as a function of time, for each initial state. It can be easily seen that only seldom the second moment of the MLL scheme results bigger than the one of the EPM scheme. It should be noted that, asymptotically the difference vanishes. The inset shows that the coherence contribution to the second moment of the EPM distribution $\langle\Delta E^2\rangle_\chi$ 
  is, in this case, negligible throughout the dynamics. The dashed black curve {corresponds to} one instance for a randomly picked initial state. We have used $\omega_k = k\omega_1$, $\gamma=0.1\omega_1$, $\beta_1=3$, $\beta_2=1$, $\beta_3=2$, $g(t)=1.5\sin^2(t)$ and $f(t)=1.5-g(t)$ with  $\hbar = k_B = 1$. }
  \label{figsec}
\end{figure}
In Fig.~\ref{figsec}, we show the difference between the second moments of the MLL and the present EPM protocol probability distributions, as a function of time and for 100 randomly chosen initial state. The inset shows that the coherence contribution is negligible. It is easy to see that, for the vast majority of cases, the second moment from the EPM protocol is greater than the one of the MLL scheme. When this happens, we can already conclude that, in the EPM protocol, we need to pay the freedom deriving from not performing any initial measurements with an increase in the uncertainty of the probability distribution. In the few cases, and instants of time, in which the hierarchy of the second moments is reversed, we need to resort to a more refined notion of uncertainty.
\begin{figure}[t]
\centering
\includegraphics[width=0.55\textwidth]{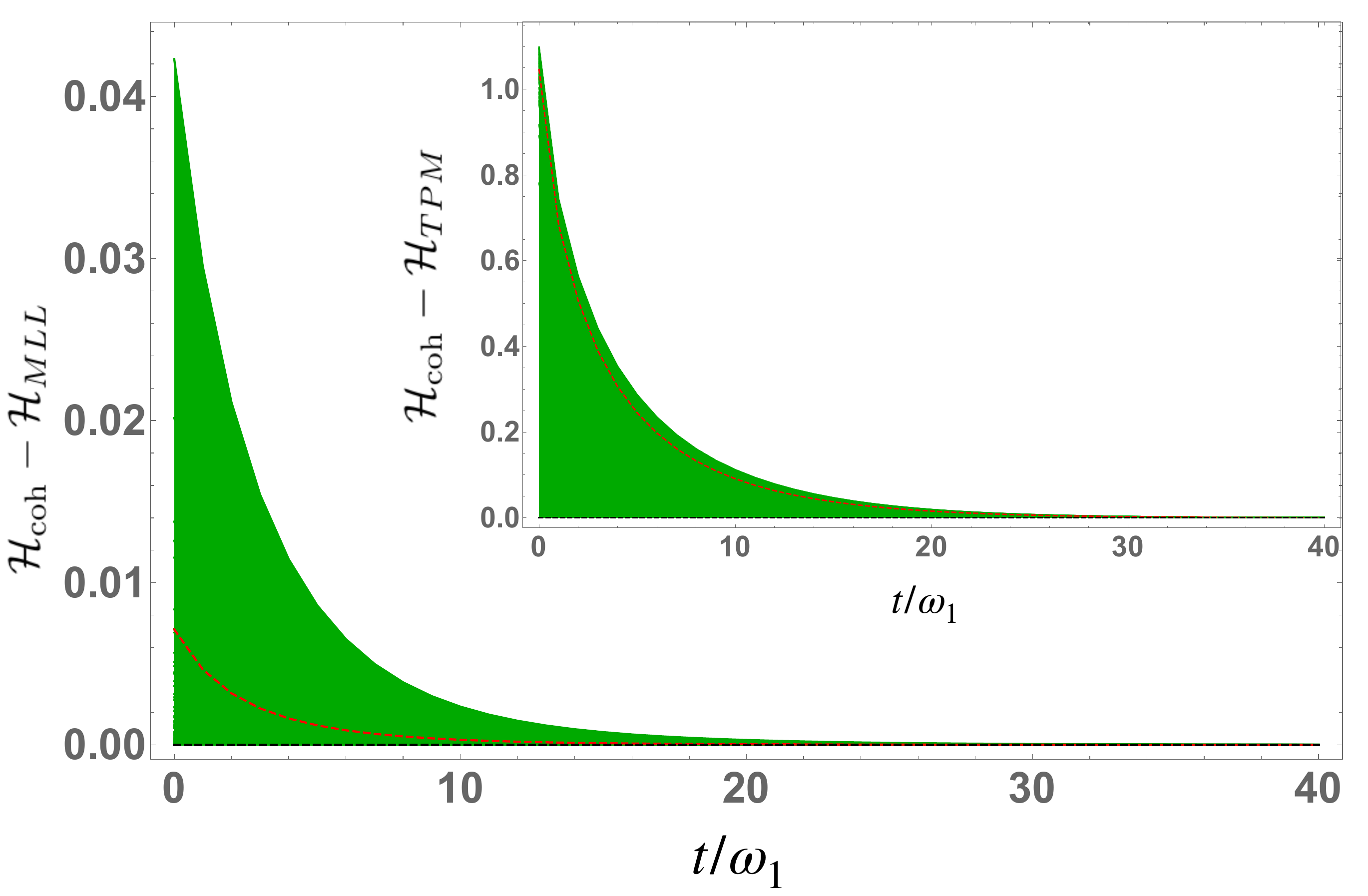}
\caption{Difference between the Shannon entropy of the EPM protocol probability distribution and the one of the MLL scheme for 100 randomly sampled initial states. The shaded area comprises all the differences, as a function of time, for each initial state. Instead, the inset shows the difference between the Shannon entropy of the EPM protocol probability distribution and the one of the TPM scheme for the same 100 randomly sampled initial states. The dashed red curve corresponds to an instance of the differences for a randomly picked initial state. It should be noted that, asymptotically, the difference vanishes (colors online).}
  \label{figentr}
\end{figure}
We do so in Fig.~\ref{figentr}, where it is shown the difference between the Shannon entropy $\mathcal{H}$ of the EPM protocol probability distribution with the one of the MLL and TPM schemes, for the same random sampling of 100 initial states as before. We see that the Shannon entropy of the EPM protocol is always greater than the one of the other schemes, which proves the increase of uncertainty due to the initial virtual measurement. While this result is expected for the comparison between the EPM and MLL schemes, as discussed before, the comparison with the TPM scheme is consistent with the fact that the effect of the initial coherence is negligible in this case. Finally, it should be noted that both the differences of second moments and Shannon entropies vanish at long times, consistently with the fact that the system reaches asymptotically a (non-equilibrium) steady-state, where all the probability distributions introduced before coincide. 
